%% file: main.tex
\documentclass[sigconf,natbib=true,anonymous=False]{acmart} 

\AtBeginDocument{%
  \providecommand\BibTeX{{%
    \normalfont B\kern-0.5em{\scshape i\kern-0.25em b}\kern-0.8em\TeX}}}
\usepackage[labelformat=simple]{subcaption}

\usepackage{enumitem}
\setlist{nosep}
\usepackage{bm}
\usepackage[linesnumbered,ruled]{algorithm2e}
\let\oldnl\nl
\newcommand{\nonl}{\renewcommand{\nl}{\let\nl\oldnl}}

\usepackage{algpseudocode}

\usepackage{booktabs}
\usepackage{float}
\usepackage{overpic}
\usepackage{float}  
\usepackage{subfloat}
\usepackage{bm}
\usepackage{multirow}
\usepackage{graphicx}
\usepackage{subcaption}
\usepackage{colortbl}
\usepackage{enumitem}
\usepackage{stfloats} 
\usepackage{url}
\usepackage{verbatim}
\usepackage{collab}
\usepackage[normalem]{ulem}
\useunder{\uline}{\ul}{}
\usepackage[skip=0.5\baselineskip]{caption}
\usepackage{color, xspace}
\usepackage{tcolorbox}  
\tcbuselibrary{breakable}   
\usepackage{amssymb}
\usepackage{pifont}
\usepackage{lipsum}
\usepackage{tcolorbox}
\usepackage{balance}

\newcommand{\name}{PAD\xspace}

\newcommand{\eat}[1]{}
\definecolor{editBlue}{RGB}{134,150,167}

\copyrightyear{2025}
\acmYear{2025}
\setcopyright{acmlicensed}\acmConference[SIGIR '25]{Proceedings of the 48th International ACM SIGIR Conference on Research and Development in Information Retrieval}{July 13--18, 2025}{Padua, Italy}
\acmBooktitle{Proceedings of the 48th International ACM SIGIR Conference on Research and Development in Information Retrieval (SIGIR '25), July 13--18, 2025, Padua, Italy}
\acmDOI{10.1145/3726302.3730059}
\acmISBN{979-8-4007-1592-1/2025/07}

\usepackage{xspace}

\newcommand{\eg}{\emph{e.g.,}\xspace}

\usepackage{marvosym}
\usepackage{ifsym}

\author{Yuhao Wang}
\affiliation{%
  \institution{City University of Hong Kong}
  \city{Hong Kong}
  \country{China}
}
\email{yhwang25-c@my.cityu.edu.hk}

\author{Junwei Pan}
\affiliation{%
  \institution{Tencent Inc.}
  \city{Shenzhen}
  \country{China}
}
\email{jonaspan@tencent.com}

\author{Pengyue Jia} 
\affiliation{%
  \institution{City University of Hong Kong}
  \city{Hong Kong}
  \country{China}
}
\email{jia.pengyue@my.cityu.edu.hk}

\author{Wanyu Wang} 
\affiliation{%
  \institution{City University of Hong Kong}
  \city{Hong Kong}
  \country{China}
}
\email{wanyuwang4-c@my.cityu.edu.hk}

\author{Maolin Wang} 
\affiliation{%
  \institution{City University of Hong Kong}
  \city{Hong Kong}
  \country{China}
}
\email{MorinWang@foxmail.com}

\author{Zhixiang Feng}
\affiliation{%
  \institution{Tencent Inc.}
  \city{Shenzhen}
  \country{China}
}
\email{lionelfeng@tencent.com}

\author{Xiaotian Li}
\affiliation{%
  \institution{Tencent Inc.}
  \city{Shenzhen}
  \country{China}
}
\email{dylanxtli@tencent.com}

\author{Jie Jiang}
\affiliation{%
  \institution{Tencent Inc.}
  \city{Shenzhen}
  \country{China}
}
\email{zeus@tencent.com}

\author{Xiangyu Zhao \Letter}
\thanks{\Letter \text{Corresponding author}}
\affiliation{%
  \institution{City University of Hong Kong}
  \city{Hong Kong}
  \country{China}
}
\email{xianzhao@cityu.edu.hk}

\renewcommand{\shortauthors}{Yuhao Wang et al.}

\begin{document}

\title{Pre-train, Align, and Disentangle: Empowering Sequential Recommendation with Large Language Models}

\begin{abstract}
Sequential Recommendation (SR) aims to leverage the sequential patterns in users' historical interactions to accurately track their preferences. However, the primary reliance of existing SR methods on collaborative data results in challenges such as the cold-start problem and sub-optimal performance. Concurrently, despite the proven effectiveness of large language models (LLMs), their integration into commercial recommender systems is impeded by issues such as high inference latency, incomplete capture of all distribution statistics, and catastrophic forgetting. To address these issues, we introduce a novel Pre-train, Align, and Disentangle (PAD) framework to enhance SR models with LLMs. In particular, we initially pre-train both the SR and LLM models to obtain collaborative and textual embeddings. Subsequently, we propose a characteristic recommendation-anchored alignment loss using multi-kernel maximum mean discrepancy with Gaussian kernels. Lastly, a triple-experts architecture, comprising aligned and modality-specific experts with disentangled embeddings, is fine-tuned in a frequency-aware manner. Experimental results on three public datasets validate the efficacy of PAD, indicating substantial enhancements and compatibility with various SR backbone models, particularly for cold items. The code and datasets are accessible for reproduction\footnote{\label{foot1}\url{https://github.com/Applied-Machine-Learning-Lab/PAD}}.
\end{abstract}


\maketitle
\input{1.introduction_new}
\input{2.method}

\input{3.experiments}

\input{4.related_work}
\input{5.conslusion}
\appendix
\input{6.appendix}

\clearpage
\bibliographystyle{ACM-Reference-Format}
\balance
\bibliography{7.reference}

\end{document}
\endinput

%% file: 1.introduction_new.tex
\section{Introduction} \label{intro}

With the explosive growth of the interactions with web applications and platforms \cite{wang2023plate,wang2024diff,zhao2018deep,wang2024llm4msr}, 
research on sequential recommender system (SRS)~\cite{SASRec2018, GRU4Rec2015, sun2019bert4rec,li2023strec,gao2024smlp4rec,zhao2022mae4rec} has garnered increasing attention. 
These models aim to capture the sequential dependencies in user behavior sequences \cite{zhang2024ssdrec,zhang2022hierarchical}, modeling both long-term and short-term preferences \cite{liu2024sequential}. 
However, most existing SR methods rely exclusively on tabular data or ID-based features as inputs \cite{zhao2018recommendations,liu2023multi,wang2023multi}, which often leads to challenges such as the cold-start problem~\cite{sheng2024enhancing}, resulting in suboptimal performance, particularly for less frequent users, items, and scenarios.

To address the limitations of conventional sequential recommendation (SR) models, recent efforts have drawn inspiration from the success of large language models (LLMs) in understanding semantics and processing natural language \cite{xu2024multi}. Several approaches have sought to enhance SR by leveraging LLM capabilities \cite{liu2024llm}. 
On the one hand, methods like TALLRec~\cite{bao2023tallrec} and LC-Rec \cite{zheng2024adapting} explicitly adapt LLMs for recommendation by instruction tuning. They formulate the sequential recommendation task as text generation, \emph{i.e.}, predicting the title of the next item based on a user's historical interactions. Afterward, LLaRA~\cite{liao2023llara}, and CoLLM~\cite{zhang2023collm} also leverage LLM as recommender and they propose to concatenating the token embedding with collaborative embedding, which enables LLM to comprehend collaborative information.

On the other hand, models such as CTRL~\cite{li2023ctrl} and Flip~\cite{wang2023flip} propose aligning LLM embeddings with collaborative embeddings using contrastive learning, employing an InfoNCE alignment loss based on non-characteristic cosine or linear kernels. Besides, Taobao \cite{sheng2024enhancing} and AlignRec \cite{liu2024aligning} propose to pre-train multi-modal representations and incorporates them into recommendation. 
Despite these advances, three practical challenges remain to be addressed:

\begin{itemize}[leftmargin=*]
    \item \textbf{High Inference Latency of LLM-based Recommenders.} LLM-based recommenders~\cite{bao2023tallrec, liao2023llara, zhang2023collm,wang2024rethinking} rely on the complex architecture of large language models to make predictions, which is impractical for real-world recommendation systems, where the inference of hundreds of items must be completed within hundreds of milliseconds~\cite{AdClickTrench2013, PracticalFacebook2014, pan2024ads}. 
    \item \textbf{Inability to Capture All Statistics of Data Distribution.} Many existing works employ non-characteristic kernels in their alignment loss~\cite{li2023ctrl, wang2023flip}. However, research on Reproducing Kernel Hilbert Space~\cite{fukumizu2008characteristic, muandet2017kernel} has demonstrated that non-characteristic kernels fail to capture all statistical aspects of the data distribution, potentially leading to suboptimal performance. 
    Evidence of the superiority of characteristic kernels in recommendation will be presented in Sec.~\ref{subsec:exp_characteristic}.
    \item \textbf{Catastrophic Forgetting in Alignment.} Catastrophic forgetting is a key challenge in multi-modal learning~\cite{li2022blip, li2023blip}. Our comprehensive empirical evaluation in Sec.~\ref{subsec:catastrophic_forgetting} shows that, although alignment helps transfer textual embeddings into the collaborative space, it results in catastrophic forgetting of the collaborative embeddings.
\end{itemize}

To address the aforementioned challenges, we propose a three-phase framework—Pre-Train, Align, and Disentangle—to empower sequential recommendation with LLMs. 
This framework consists of (1) pre-training LLM and recommendation models, (2) alignment between textual and collaborative with MK-MMD, and (3) supervised fine-tuning for recommendation. Specifically, during the pre-training phase, we employ a pre-trained LLM, \eg Llama3~\cite{dubey2024llama}, alongside a sequential recommendation model, \eg SASRec~\cite{SASRec2018}. 
The textual and collaborative embeddings are then obtained from these two models, respectively.

In the alignment phase, we propose a novel characteristic recommendation anchored alignment loss, which integrates a characteristic alignment loss with a Binary Cross Entropy (BCE) loss based on the recommendation label. The characteristic alignment loss ensures that all statistical properties of the distribution are accounted for during alignment, while the BCE loss mitigates catastrophic forgetting of the collaborative embeddings during the alignment.

In the subsequent disentangle phase, inspired by recent advances in recommendation for handling the distinct characteristics of tasks~\cite{su2024stem}, domains~\cite{lin2024disentangled}, and modalities~\cite{sheng2024enhancing}, we incorporate two modality-specific embeddings and experts, in addition to the alignment expert, resulting in a triple-expert architecture. Specifically, we employ an LLM expert and a recommendation expert, which use the LLM embeddings and collaborative embeddings as inputs, respectively. 
The key contributions of this paper are as follows:

\begin{itemize}[leftmargin=*]
    \item We identify the limitations of existing SRS works in capturing distribution statistics and catastrophic forgetting. 
    \item We propose a novel three-phase framework—Pre-train, Align, and Disentangle (\name)—featuring a recommendation-anchored characteristic alignment loss and a triple-expert architecture.
    \item We conduct extensive experiments on three public datasets, demonstrating the effectiveness of \name, especially on cold items. 
    Furthermore, we provide a comprehensive study regarding the kernels, catastrophic forgetting and tools to measure alignment. 
    
\end{itemize}

%% file: 2.method.tex
\vspace{-0.3em}
\section{Preliminary} \label{prel}
In this section, we first illustrate the problem formulation and introduce maximum mean discrepancy.

\vspace{-0.3em}
\subsection{Problem Formulation}

First we provide the formulation and notations. In our problem setting, we obtain a semantic domain $\mathcal{D}_\text{text} = \{ (\{\mathbf{h}_i^s\},$ $\mathbf{x}_i^s,y_i) \}_{i=1}^{n}$ and a collaborative domain $\mathcal{D}_\text{rec} = \{ (\{\mathbf{h}_i^c\},\mathbf{x}_i^c,y_i) \}_{i=1}^{n}$ with $n$ samples each. Specifically, $\{\mathbf{h}_i^s\}$, $\{\mathbf{h}_i^c\}$, $\mathbf{x}_i^s$, $\mathbf{x}_i^c$, and $y_i$ denotes behavioral item sequence in semantic embedding space, behavioral item sequence in collaborative embedding space, target item in semantic embedding space, target item in collaborative embedding space, and true label. The probability distributions characterized by these two domains are $P$ and $Q$.

Given a user's historical interaction sequence with length $l$ consisting of item ID, it is first sorted by timestamps in an ascending order and mapped into collaborative embedding $\{\mathbf{h}_i^c\}, i=1 \ldots l$. Then sequential recommender system (SRS) $f_ \theta$ usually takes it as input and outputs user embedding, which is multiplied by target item embedding $\mathbf{x}_i^c$ through dot product to obtain the prediction logit. Finally, the binary cross entropy (BCE) loss is often adopted.
\begin{equation}
\label{bce}
\setlength{\abovedisplayskip}{3pt}
\setlength{\belowdisplayskip}{1pt}
\min _{\theta} \mathcal{L}=\frac{1}{n} \sum_{i=1}^{n} \text{BCE}\left(f_ \theta\left( 
\{\mathbf{h}_i^c\}, \mathbf{x}_i^c \right), y_i\right)
\end{equation}



\subsection{Maximum Mean Discrepancy} \label{pre-mmd}

Kernel function $k(\cdot,\cdot)$ characterizes how to measure similarity between samples.
The idea of \textit{kernel mean} or \textit{mean embedding} is to represent the distribution in the reproducing kernel Hilbert space (RKHS)~\cite{muandet2017kernel}.
Specifically, considering a symmetric, positive-definite kernel $k$ and its corresponding unique RKHS $\mathcal{H}_k$, each distribution $P(X)$ is mapped into $\mathcal{H}_k$ through $\mu_P \triangleq \mathbb{E}[k(X,\cdot)]=\mathbb{E}[\varphi(X)]$ where $\varphi$ is the feature map.



Furthermore, the idea of Maximum Mean Discrepancy (MMD) is to represent the distance between distributions as the distance between kernel mean of features~\cite{sejdinovic2013equivalence}. 
One can define the distance between probability distributions $P$ and $Q$ in RKHS $\mathcal{H}_k$ as: 
\begin{equation}
\setlength{\abovedisplayskip}{3pt}
\setlength{\belowdisplayskip}{1pt}
D_k(P, Q) \triangleq \| \mu_P - \mu_Q \|_{\mathcal{H}_k}, 
\end{equation}
where $\| \cdot \|_{\mathcal{H}_k}$ is the norm of $\mathcal{H}_k$ and this metric is known as MMD~\cite{sejdinovic2013equivalence}.
As illustrated by previous works~\cite{gretton2012optimal}, the choice of the kernel will affect the power of the two-sample test, whose null hypothesis is $P=Q$.
If the positive definite kernel $k$ is characteristic, \emph{i.e.}, the mapping $ P \mapsto \mu_{P} \in \mathcal{H}_k $ is injective, the kernel mean is proven to preserve all information of the distribution $P(X)$~\cite{fukumizu2004dimensionality}.
Besides, the following multi-kernel MMD (MK-MMD)~\cite{gretton2012kernel} is introduced to improve test power where
$k$ is a kernel function in the function space. Formally,
\begin{equation}
\label{space}
\setlength{\abovedisplayskip}{3pt}
\setlength{\belowdisplayskip}{1pt}
\mathcal{K} \triangleq \{ k=\sum_{u=1}^{m}\beta_{u} k_{u}, \sum_{u=1}^{m}\beta_{u}=d, \beta_{u} \geq 0, \forall u \}
\end{equation}
for some $d \ge 0$ where $\{k_{u}\}_{u=1}^{m}$ are positive definite single kernels.

\section{Method}

In this section, we first briefly present the three phases in Sec.~\ref{subsec:overview}, and then provide a detailed description of the align and disentangle phase in Sec.~\ref{subsec:alignment} and Sec.~\ref{subsec:finetune}, respectively.


\subsection{Overview} \label{subsec:overview}

\begin{figure*}[t]
    \centering
    \setlength\abovecaptionskip{0.2\baselineskip}
    \setlength\belowcaptionskip{0.2\baselineskip}
    \includegraphics[scale=0.445, trim=0 5 8 0,clip]{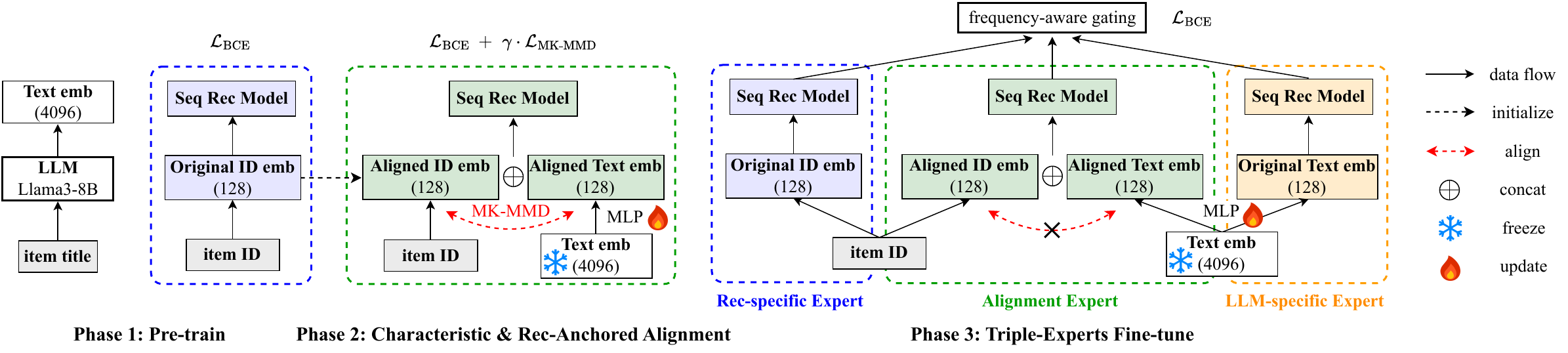}
    \caption{Overall framework of \name. The number in parentheses (128 and 4096) denotes the embedding dimension. The prediction logit is calculated by multiplying the sequence embedding (output of recommendation model) and target item embedding. For simplicity the multiplication operation is omitted.}
    \label{framework}
    \vspace{-3mm}
\end{figure*}

The overview process of our model is depicted in Fig.~\ref{framework}, which consists of the following three phases: 
\vspace{-0.3em}
\paragraph{\textbf{Phase 1: LLM \& Recommendation Pre-train}}
First, the textual embedding is generated from the textual information of items like titles and descriptions, and then it is frozen as fixed semantic knowledge. 
We adopt LLM2Vec~\cite{behnamghader2024llm2vec} which transforms the primary large language models (LLMs) with decoder-only structure like Llama3 into powerful text encoders through fine-tuning.
Next, the recommendation (ID) expert is pre-trained only on item ID to capture collaborative information using a SASRec model. 

\vspace{-0.3em}
\paragraph{\textbf{Phase 2: Characteristic \& Rec-Anchored Alignment}}
After obtaining the textual and collaborative embeddings from the pre-trained LLM and recommendation model, we will align these two embeddings with each training sample.
In particular, we build an alignment expert, which takes both textual and collaborative embedding as inputs, and adopts MK-MMD~\cite{fukumizu2004dimensionality} as the alignment loss with characteristic kernels, thus being able to preserve all information about the distribution~\cite{fukumizu2008characteristic}.
In addition, we introduce a Binary Cross Entropy loss regarding the recommendation label so as to present the collaborative embeddings from catastrophic forgetting~\cite{li2023blip} during alignment.




\vspace{-0.3em}
\paragraph{\textbf{Phase 3: Collaborative Fine-tuning}}
In this phase, besides the alignment expert, we introduce two modality-specific experts, \emph{i.e.}, one takes only the textual embedding as input, while the other takes the pre-trained collaborative embedding as input.
These three experts are combined via a Mixture-of-Experts (MoE) architecture through a frequency-aware gating mechanism and then fine-tuned by the collaborative supervision signals.

\subsection{Characteristic \& Rec-Anchored Alignment} \label{subsec:alignment}

Existing methods~\cite{li2023ctrl} mainly adopt non-characteristic kernels and alignment loss is usually the only optimization objective.
Nonetheless, they would fail to grasp all statistics of the data distribution and suffer catastrophic forgetting which will be detailed in Sec.~\ref{subsec:exp_rec_anchor}.
Therefore, to tackle these limitations, we employ an alignment expert to align the textual embeddings towards the collaborative space.
Specifically, we adopt MK-MMD~\cite{sejdinovic2013equivalence, gretton2012optimal} with characteristic kernels as the alignment loss since it's able to capture all information about the distribution~\cite{fukumizu2004dimensionality, muandet2017kernel}.
Besides, we introduce an auxiliary recommendation Binary Cross Entropy loss to conduct Rec-Anchored alignment. It could enable the collaborative embeddings keep as much recommendation semantics as possible after the alignment, also avoiding catastrophic forgetting.
The collaborative embeddings are not frozen during the alignment stage.
By contrast, the pre-trained text embeddings remain frozen while the MLPs for dimension reduction are learnable.
The overall loss function is:
\begin{align}
    \label{loss}
    \setlength{\abovedisplayskip}{3pt}
    \setlength{\belowdisplayskip}{1pt}
    &\mathcal{L} = \mathcal{L}_\text{REC} + \gamma \cdot \mathcal{L}_\text{MK-MMD} \\
    &\mathcal{L}_\text{MK-MMD} =  D_k^2\left(\{\mathbf{h}_i^s\}_\text{a}, \{\mathbf{h}_i^c\}_\text{a}\right) \\
     &\mathcal{L}_\text{REC} = \frac{1}{n} \sum_{i=1}^{n} \text{BCE}\left(f_ \theta\left( 
    \{\mathbf{h}_i^s\}_\text{a}, \{\mathbf{h}_i^c\}_\text{a}, {\mathbf{x}_i^s}_\text{a}, {\mathbf{x}_i^c}_\text{a} \right), y_i\right) \\
    & \{\mathbf{h}_i^s\}_\text{a} = f_\text{MLP}(\texttt{SG}(\{h_\text{text}\}), \bm{w})
\end{align}

\noindent where $\gamma$ is the hyper-parameter.  The subscript $a$ denotes the newly aligned embedding.
$k$ is the combination of multiple characteristic kernels. $f_\theta$ is the recommender system taking both collaborative and text data as input. $\{h_\text{text}\}$ denotes the textual embedding from LLM2Vec. $\texttt{SG}$ denotes the stop gradient operation on the textual embedding. $f_\text{MLP}$ is a learnable fully connected network with parameter $\bm{w}$ reducing the dimension of textual embedding. Other notations have been illustrated at the beginning of Sec.~\ref{prel}.

\subsection{Collaborative Fine-tuning}\label{subsec:finetune}


Existing works~\cite{du2024disco,yang2024darec} simply append the aligned embeddings to the input of recommendation model and regard it as additional features. Nevertheless, these embeddings may not be fully comprehended and exploited by the model.
Therefore, in this stage, in order to fully utilize the aligned representation to enhance the performance of recommendation tasks,
we propose a triple-experts architecture with four embedding tables (\emph{i.e.}, two for each modality). Specifically, it consists of an alignment expert, an LLM-specific expert, and a recommendation-specific expert.
Besides, existing works \cite{zhang2023collm,liao2023llara} neglect the impact of item frequency on the credibility of different modality information. 
Therefore, we propose to fuse the output of these three experts via a frequency-aware gating.

\vspace{-0.3em}
\paragraph{\textbf{Triple-Experts Architecture}}
To fully exploit different modality data and alleviate catastrophic
forgetting, our model consists of an aligned expert $f_\text{align}$, which takes the aligned  $\{\mathbf{h}_i^s\}_\text{a}$ and $\{\mathbf{h}_i^c\}_\text{a}$ as input, and two modality-specific experts $f_\text{LLM}$ and $f_\text{id}$, which take only the original textual embedding $\{\mathbf{h}_i^s\}$ or pre-trained collaborative embedding $\{\mathbf{h}_i^c\}$ as input.
The MK-MMD alignment loss is removed in the alignment expert in this phase and  $\mathcal{L}_\text{REC}$ only is the loss function.
Besides, the textual embeddings in both the alignment and LLM-specific experts are frozen, while the collaborative embeddings in the alignment and recommendation-specific experts are being updated. Finally, the output of each expert is obtained:
\begin{align}
    \label{output}
    \setlength{\abovedisplayskip}{1pt}
    \setlength{\belowdisplayskip}{1pt}
    &o_\text{id} = f_\text{id}(\{\mathbf{h}_i^c\}) \\
    &o_\text{align} = f_\text{align}(f_\text{MLP}(\texttt{SG}(\{\mathbf{h}_i^s\}_\text{a}), \bm{w}), \{\mathbf{h}_i^c\}_\text{a}) \\
     &o_\text{LLM} = f_\text{LLM}(f_\text{MLP}(\texttt{SG}(\{\mathbf{h}_i^s\}), \bm{w}^{\prime}))
\end{align}

\noindent Such a design shares spirits with the Multi-Embedding paradigm~\cite{guo2023embedding, su2024stem,lin2024disentangled} in the sense that for each modality we have two embedding tables, one for the alignment expert and the other one for the recommendation- or LLM-specific expert.

\vspace{-0.3em}
\paragraph{\textbf{Frequency-aware Gating}} We propose to fuse the output of the three experts based on the target item frequency in an adaptive manner. To be specific, we first divide all items into $B$ buckets based on their frequency. Next, given the target item $i$ and its corresponding bucket ID $b(i)$, a gating network $g$ is learned which takes $b(i)$ and the expert embedding as input to generate the probability for each expert. 
Finally, we fuse the output of each expert with a frequency-aware gating mechanism to generate the final logit:
\begin{align}
\label{fusion}
    \setlength{\abovedisplayskip}{3pt}
    \setlength{\belowdisplayskip}{1pt}
    \Phi &= g_\text{id}(b(i),\{\mathbf{h}_i^c\}) \cdot o_\text{id} \nonumber \\
    &+ g_\text{align}(b(i),\{\mathbf{h}_i^s\}_\text{a},\{\mathbf{h}_i^c\}_\text{a}) \cdot o_\text{align} \nonumber \\
     &+ g_\text{LLM}(b(i),\{\mathbf{h}_i^s\}) \cdot o_\text{LLM}
\end{align}

\noindent where $g_\text{id}$, $g_\text{align}$, and $g_\text{LLM}$ denote the probability of recommendation-specific, alignment, and LLM-specific expert, respectively.

%% file: 3.experiments.tex
\section{Experiments} \label{expe}

We conduct extensive experiments on three public datasets and answer the following six research questions:
\begin{itemize}[leftmargin=*]
\item \textbf{RQ1:} How does \name perform, compared with other SOTA baseline methods?
\item \textbf{RQ2: } What is the effect of recommendation anchoring in the alignment loss?
\item \textbf{RQ3: } How do the characteristic alignment losses perform, compared with non-characteristic ones?
\item \textbf{RQ4: } How do the modality-specific embedding and experts contribute to the performance enhancement?
\item \textbf{RQ5:} What is the impact of each expert (ID, alignment, and LLM-specific expert) and the proposed frequency-aware fusion?
\item \textbf{RQ6:} Do large language models (LLMs) have more powerful encoding capability than pre-trained language models (PLMs)?
\item \textbf{RQ7:}  Is \name compatible with other sequential recommendation models as a model-agnostic paradigm?


\end{itemize}

\subsection{Experimental Settings}
\subsubsection{\textbf{Datasets}} Our experiments are conducted on three datasets: MIND~\cite{wu2020mind}, Electronics, and Prime Pantry where the last two are two categories from Amazon~\cite{ni2019justifying}. Their statistics are summarized in Tab.~\ref{tab:stat}. The detailed data preprocess and split procedure is demonstrated in Appendix.~\ref{detail}. 

\begin{itemize}[leftmargin=*]
\vspace{1mm}
\item \textbf{MIND\footnote{\url{https://msnews.github.io/}}} is collected from the Microsoft news website. It contains abundant text information with news title, abstract, body, etc. Given the historical click events of user, the task is to predict whether this user would click the target news.

\item \textbf{Amazon\footnote{\url{https://cseweb.ucsd.edu/\textasciitilde jmcauley/datasets/amazon\_v2/}}} is collected from the e-commerce platform Amazon in which users rate items from 1 to 5. The sequential recommendation task is to predict whether a user will give a rating higher than 3 to the target item.

\end{itemize}

\subsubsection{\textbf{Evaluation Metrics}} 
We choose the widely-used top-k Hit Ratio (HR@k) and top-k normalized Discounted Cumulative
Gain (nDCG@k) with k = 10 for evaluation on the
whole item set. The values reported below are averaged over all users.

\subsubsection{\textbf{Baselines}} \label{baseline}
The following representative baseline methods are chosen for comparison. 



\begin{table}[t] \tiny
\centering
\setlength\abovecaptionskip{0.3\baselineskip}
\setlength\belowcaptionskip{0.3\baselineskip}
\caption{The statistics of three public datasets: MIND, Electronics, and Prime Pantry category of Amazon.}
    \label{tab:stat}
    \resizebox{0.4\textwidth}{!}{
    \setlength{\tabcolsep}{1mm}{
    \begin{tabular}{cccccc}
    \toprule[0.7pt]
        Dataset & \multicolumn{1}{c}{Users} & \multicolumn{1}{c}{Items} & Interactions \\
    \midrule    
        MIND & 630,235 & 79,707 & 9,667,540  \\
        Electronics & 598,307 & 423,191 & 5,137,265 \\
        Prime Pantry & 12,871 & 8,347 & 115,004 \\
    \bottomrule[0.7pt]
    \end{tabular}}
    \vspace{3mm}
    }
\end{table}


\begin{itemize}[leftmargin=*]
\item \textbf{SASRec}~\cite{kang2018self}
denotes the primary SASRec model, which is only trained by pure ID of items in the collaborative space.


\item \textbf{Hybrid} denotes Hybrid Encoding adopted in CoLLM~\cite{zhang2023collm} and LLaRA~\cite{liao2023llara}. It directly concatenates the collaborative with text embedding and feeds into the subsequent SRS.
\item \textbf{MoRec}~\cite{yuan2023go} simply adopts an MLP to reduce the dimension of text embedding and inputs it into SRS. 
\item \textbf{CTRL}~\cite{li2023ctrl} first pre-trains the model parameters by aligning the collaborative with text embedding using contrastive learning, then fine-tunes on the recommendation task.
\item \textbf{MAKE}~\cite{sheng2024enhancing} adopts a two-experts structure. It first pre-trains the LLM-specific expert in the same way as MoRec, then fuse the output of ID and LLM-specific expert to generate prediction.
\item \textbf{DisCo}~\cite{du2024disco} splits the embeddings into chunks and incorporates sufficiency and disentanglement loss to explicitly preserve the task-related and unique information.
\item \textbf{SMEM} is a novel baseline we propose, which adopts Shared
and Modality-specific EMbeddings. It is similar to Shared
and Task-specific EMbeddings (STEM)~\cite{su2024stem} in multi-task recommendation.
\end{itemize}



\subsubsection{\textbf{Implementation Details.}}
We implement \name with multiple Gaussian kernels: 
\begin{equation}
    \setlength{\abovedisplayskip}{1pt}
    \setlength{\belowdisplayskip}{1pt}
k_u(x,x') = \text{exp} ( -\frac{\|x-x'\|^2}{2\sigma^2})
\end{equation}
where $k_u \in \mathcal{K}$ in \eqref{space}. We take $m=5$ and $\sigma=\{ -3,-2,-1,0,1\}$ in the following experiments.
Besides, our experiments are conducted on Tesla V100 GPUs and all the results shown are averaged over 3 runs. Detailed experimental settings are provided in Appendix.~\ref{detail}.

\subsection{Overall Performance (RQ1)}

\begin{table*}[ht!]

\centering
\setlength\abovecaptionskip{0.2\baselineskip}
\setlength\belowcaptionskip{0\baselineskip}
\caption{Overall performance comparison on MIND, Electronics, and Prime Pantry dataset. Boldface denotes the highest value while underline indicates the second best result. `Impr.' indicates our improvement against the original SASRec. $\star$ represents statistical significance with $p$-value $< 0.05$ in $t$-test compared with the best baseline.}
\label{tab:overall}

\resizebox{0.86\textwidth}{!}{
\begin{tabular}{@{}cccccccccccc@{}}
\toprule
\multicolumn{1}{c}{Datasets} & \multicolumn{1}{c}{Metric} & \multicolumn{1}{c}{SASRec} & \multicolumn{1}{c}{Hybrid} & \multicolumn{1}{c}{MoRec} & \multicolumn{1}{c}{CTRL} & \multicolumn{1}{c}{MAKE} & \multicolumn{1}{c}{DisCo} & \multicolumn{1}{c}{SMEM} & \multicolumn{1}{c}{Ours} & \multicolumn{1}{c}{Impr.} \\ \midrule
\multirow{2}{*}{{MIND}} & HR@10 & 16.8437 & 17.3385 & 14.1235 & 18.1318 & 16.4317 & 17.3367 & {\ul 18.4319} & \bm{$18.6703^{\star}$} & 10.84\% \\ 
 & nDCG@10 & 9.0520 & 9.3321 & 7.7090 &  9.7996  & 8.9684 & 9.4131 & {\ul 10.0003} & \bm{$10.1515^{\star}$} & 12.15\% \\ \midrule 
\multirow{2}{*}{{Electronics}} & HR@10 & 1.6754 & 1.5673 & 0.9562  & 1.9468 & 0.9495 & 1.8148 & {\ul 2.3492} & \bm{$2.4804^{\star}$} & 48.05\% \\ 
 & nDCG@10 & 0.7938 & 0.8385 & 0.4721 & 0.9370  & 0.4611 & 0.9025 & {\ul 1.4287} & \bm{$1.5373^{\star}$} & 93.67\% \\ \midrule 
\multirow{2}{*}{{Prime Pantry}} & HR@10 & 2.6338 & 3.1000 & 3.1699  & 2.8048 & 3.3408 & 2.8358 & {\ul 3.4341} & \bm{$3.8303^{\star}$} & 45.43\% \\
 & nDCG@10 & 1.2926 & 1.5606 & 1.6013 & 1.3700 & 1.6379 & 1.3451 & {\ul 1.7440} & \bm{$1.9104^{\star}$} & 47.80\% \\

\bottomrule
\end{tabular}}

\vspace{-3mm}
\end{table*}

To answer \textbf{RQ1}, we verify \name's effectiveness by comparing it with various baseline methods introduced in Sec.~\ref{baseline}. 
The overall performance on three public datasets is shown in Tab.~\ref{tab:overall}. 
In summary, \emph{on all three datasets, our proposed \name achieves state-of-the-art performance, beating the best performing baseline SMEM by 1.51\%, 7.60\%, and 9.54\%  on nDCG@10 on each dataset}.

\begin{figure}[t]
\setlength\abovecaptionskip{-0.8\baselineskip}
\setlength\belowcaptionskip{0.6\baselineskip}
	\centering
        \begin{minipage}{0.32\linewidth}
		\centering
        \begin{subfigure}{1\linewidth}
		\includegraphics[width=0.995\linewidth]{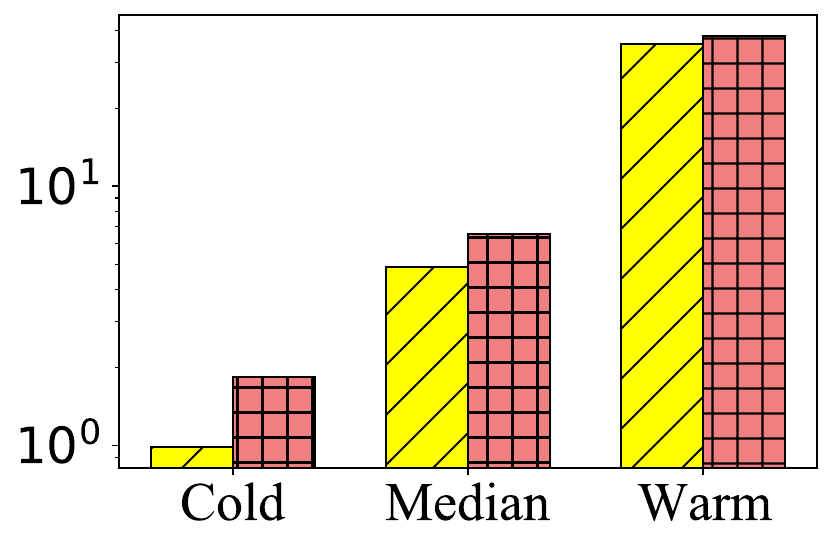}
		\label{fig:RQ3-4}
        \end{subfigure}
	\end{minipage}
	\begin{minipage}{0.32\linewidth}
		\centering
        \begin{subfigure}{1\linewidth}
		\includegraphics[width=0.995\linewidth]{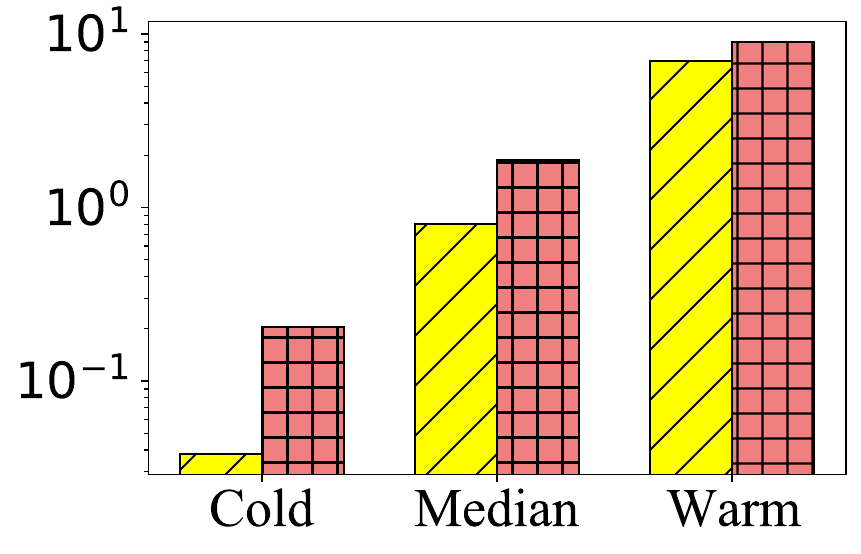}
		\label{fig:RQ3-1}
        \end{subfigure}
	\end{minipage}
	\begin{minipage}{0.32\linewidth}
		\centering
        \begin{subfigure}{1\linewidth}
		\includegraphics[width=0.995\linewidth]{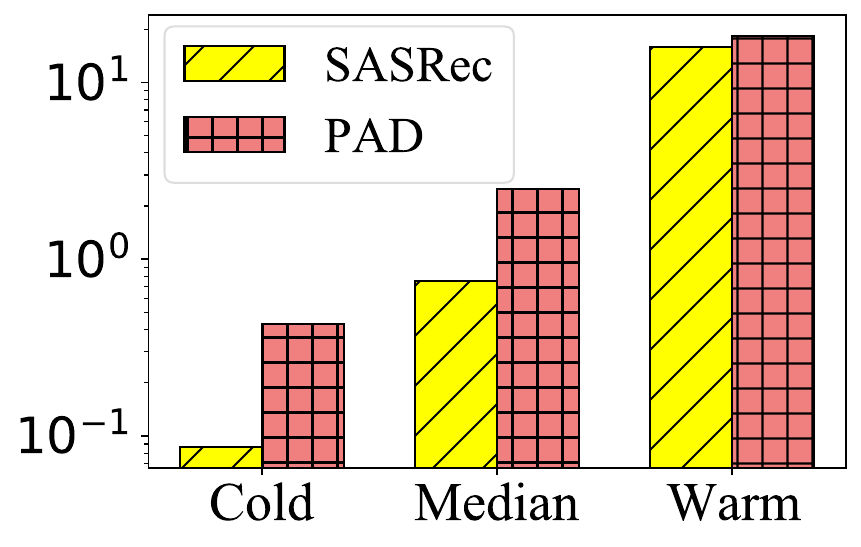}
		\label{fig:RQ3-2}
        \end{subfigure}
	\end{minipage}
	\caption{Comparison of the original SASRec and \name on the Mind (left), Electronics (mid) and Prime Pantry (right) datasets. Warm, median, cold denote target items with high to low frequency on the test set. The y-axis denotes the HR@10.} 
	\label{fig:bucket}
\end{figure}

Moreover, in Fig.~\ref{fig:bucket} we also provide the results on different subsets of items, \emph{i.e.}, on warm, median, and cold items. 
We observe that \name surpasses SASRec on all the item subsets on each dataset, and the improvement on cold items is more significant.
For example, on Electronics dataset, our method achieves 63.36\%, 202.02\% and 462.61\% relative performance lift on nDCG@10 on the warm, median and cold items.
The performance lift on the cold items are 7.3 times of that on the warm items.
\emph{This validates that our method can mitigate the cold-start problem with the LLM knowledge effectively}.

\begin{figure*}[t]
\setlength\abovecaptionskip{-1\baselineskip}
\setlength\belowcaptionskip{-0.2\baselineskip}
	\centering
        \begin{minipage}{0.47\linewidth}
		\centering
        \begin{subfigure}{1\linewidth}
		\includegraphics[width=0.995\linewidth]{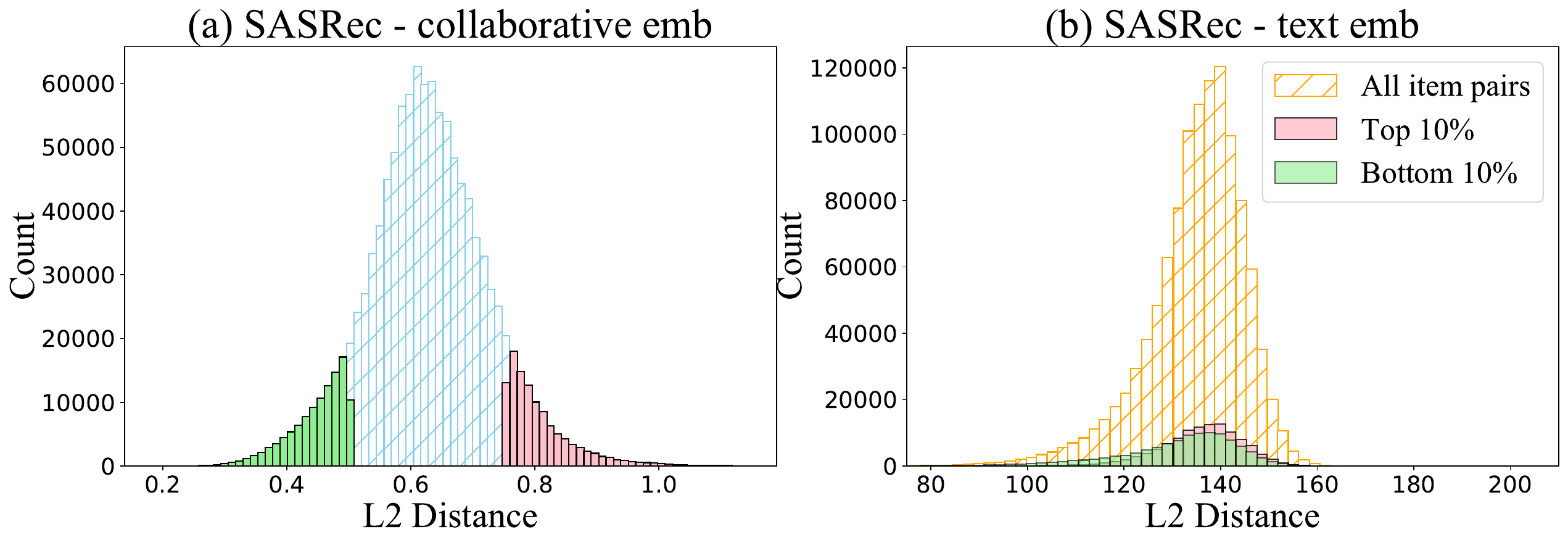}
		\label{fig:RQ3-4}
        \end{subfigure}
	\end{minipage}
        \begin{minipage}{0.47\linewidth}
		\centering
        \begin{subfigure}{1\linewidth}
		\includegraphics[width=0.995\linewidth]{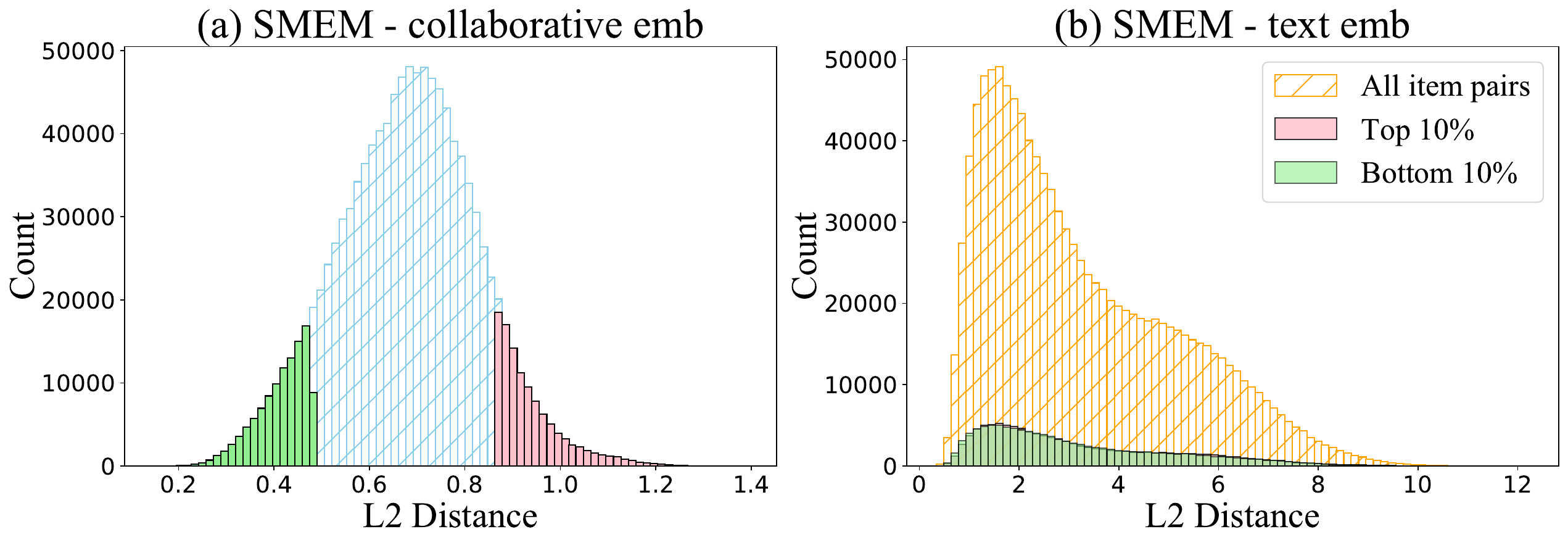}
		\label{fig:RQ3-4}
        \end{subfigure}
	\end{minipage}

        \vspace{-2mm}
 
        \begin{minipage}{0.47\linewidth}
		\centering
        \begin{subfigure}{1\linewidth}
		\includegraphics[width=0.995\linewidth]{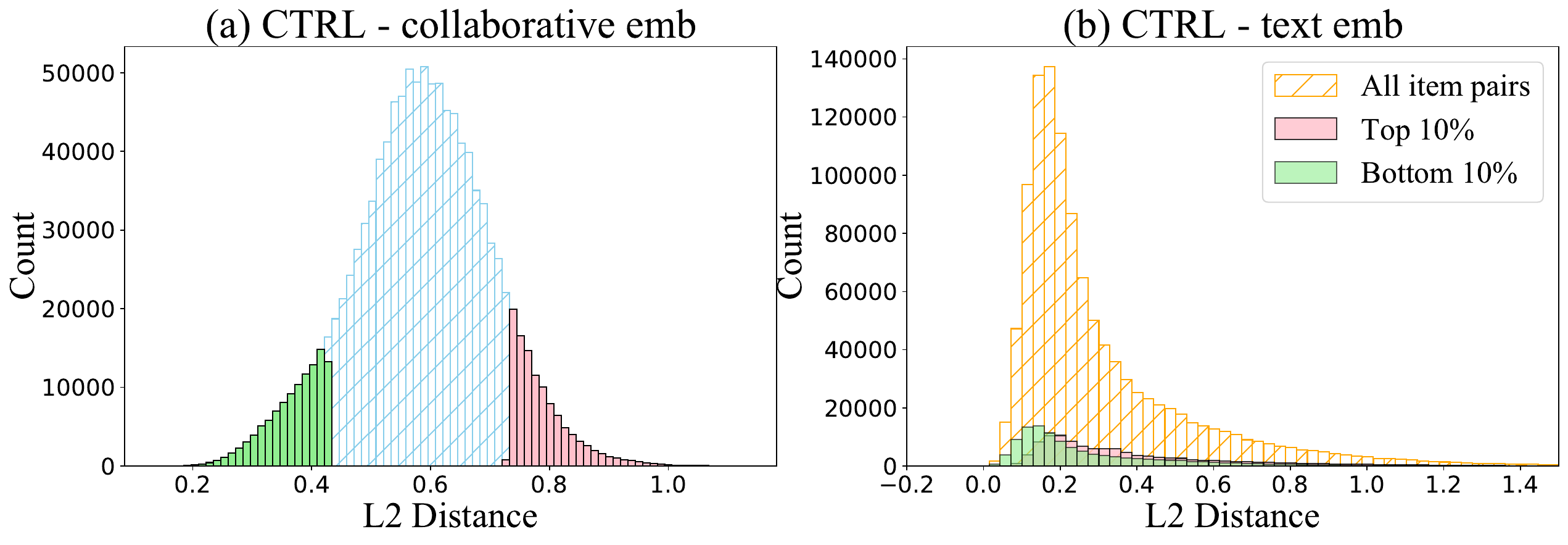}
		\label{fig:RQ3-4}
        \end{subfigure}
	\end{minipage}
        \begin{minipage}{0.47\linewidth}
		\centering
        \begin{subfigure}{1\linewidth}
		\includegraphics[width=0.995\linewidth]{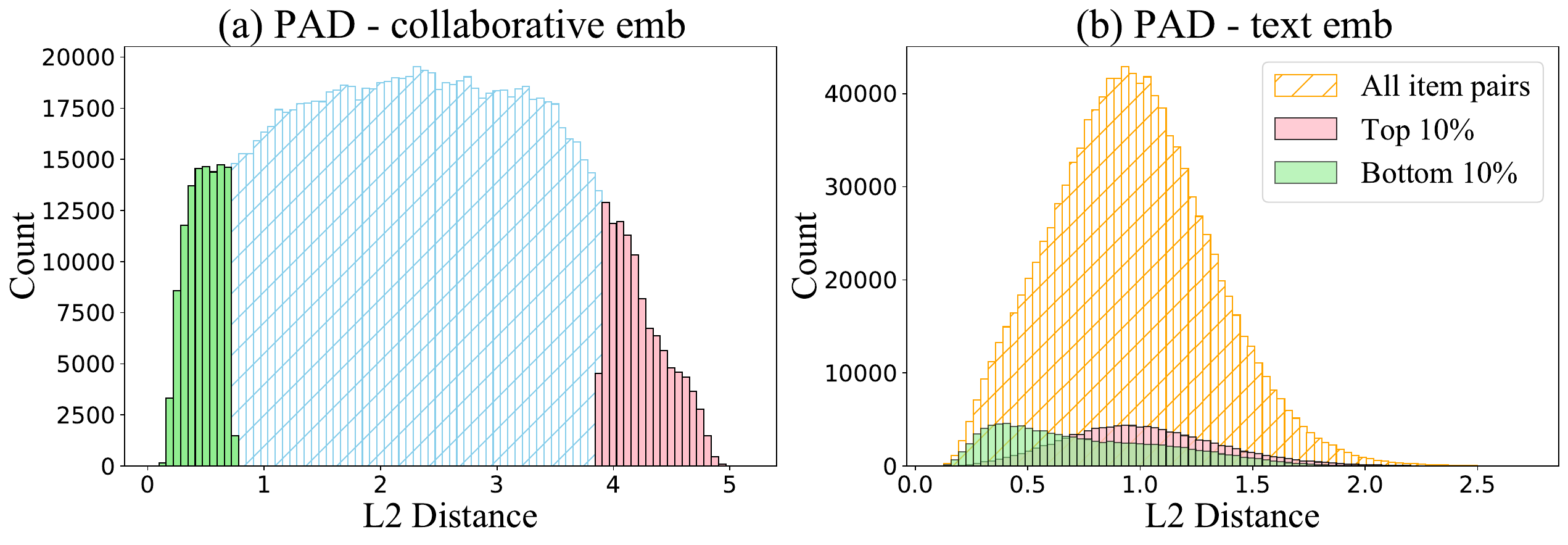}
		\label{fig:RQ3-4}
        \end{subfigure}
	\end{minipage}
	\caption{$\mathcal{P}_{\text{Top-10\% }}^\text{ID}$ and $\mathcal{P}_{\text{Bottom-10\% }}^\text{ID}$ under the distance distribution regarding the collaborative and textual embeddings in original SASRec, SMEM, CTRL, and our proposed \name on cold items.} 
	\label{app:cold}
\vspace{-2mm}
\end{figure*}

We further illustrate the effectiveness of our method on the cold items with the following analysis.
Denote $\mathcal{P}_{\text{Top-10\% }}^\text{ID}$ and $\mathcal{P}_{\text{Bottom-10\% }}^\text{ID}$ as the group of item pairs with the top-10\% and bottom-10\% distance based on collaborative collaborative embeddings. 
We present the distribution of these item pairs under the distance distribution regarding the collaborative embeddings and textual embeddings in each model in Fig.~\ref{app:cold}.
We observe that all existing methods, including SASRec, SMEM, and CTRL, can not differentiate the distribution of $\mathcal{P}_{\text{Top-10\% }}^\text{ID}$ and $\mathcal{P}_{\text{Bottom-10\% }}^\text{ID}$ well.
In contrast, \emph{our method succeeds in learning a generally larger textual distance for those item pairs with top collaborative distance and learning a smaller textual distance for those with bottom collaborative distance}.
Consequently, we conclude with the following result:

\begin{tcolorbox}[colback=blue!2!white,leftrule=2.5mm,size=title]
    \emph{Result 1. Our proposed \name greatly improves the performance on the cold items with better alignment of textual embeddings towards the collaborative space, leading to SOTA performance on three datasets. }
\end{tcolorbox}

\subsection{Discrepancy Measurement}
\label{subsec:discrepancy_measurement}

In this section, we propose a novel metric to quantify to what extent are the aligned embedding space deviates from the original one.
Specifically, rather than the deviation of the aligned and original embedding for each ID itself, \emph{we are more interested in the discrepancy of embedding distance}~\cite{su2024stem}.
To this end, we employ the Kendall's tau~\cite{kendall1938new} between the embedding distance distribution of the original embeddings and that of the aligned embeddings.

Specifically, in sequential recommendation, the embedding distance between each behavior and target item pair is calculated.
Afterward, Kendall's tau is adopted to measure the degree of concordance between these two variables. 
For example, given the user's historical interaction sequence $\{i_1,i_2,i_3\}$, the SRS usually takes $\{i_1\}$ as the behavior item sequence to predict the target item $i_2$ and takes $\{i_1,i_2\}$ as the behavior item sequence to predict the target item $i_3$. We map each item $i_s$ into embedding $\bm{e}_s$ and calculate the distance $\langle \cdot~,\cdot \rangle$ (like Euclidean distance) between each behavior and target item embedding pair, which includes $\{ \langle \bm{e}_1,\bm{e}_2 \rangle, \langle \bm{e}_1,\bm{e}_3 \rangle, \langle \bm{e}_2,\bm{e}_3 \rangle \}$. 
For another model with embeddings $\bm{e}'_s$ for $i_s$, its distance variable is $\{ \langle \bm{e}'_1,\bm{e}'_2 \rangle, \langle \bm{e}'_1,\bm{e}'_3 \rangle, \langle \bm{e}'_2,\bm{e}'_3 \rangle \}$. 
Any pair of samples is called concordant if the sort order agrees, \emph{e.g.}, both $\langle \bm{e}_1,\bm{e}_2 \rangle < \langle \bm{e}_1,\bm{e}_3 \rangle$ and $\langle \bm{e}'_1,\bm{e}'_2 \rangle < \langle \bm{e}'_1,\bm{e}'_3 \rangle$ hold or $\langle \bm{e}_1,\bm{e}_2 \rangle > \langle \bm{e}_1,\bm{e}_3 \rangle$ and $\langle \bm{e}'_1,\bm{e}'_2 \rangle > \langle \bm{e}'_1,\bm{e}'_3 \rangle$ hold. The Kendall's tau is defined as
\begin{equation}
    \setlength{\abovedisplayskip}{3pt}
    \setlength{\belowdisplayskip}{1pt}
    \text{KT} = \frac{\#(\text{concordant pairs})-\#(\text{disconcordant pairs})}{\#(\text{pairs})}
\end{equation}
where $\#$ denotes the number.
Notably, this method of measurement is also applicable to the circumstance where the dimensions of embedding are inconsistent between models or modalities, and it can be easily extended to other binary relation like user-user and user-item distance.

\begin{figure}[t]
\setlength\abovecaptionskip{-0.8\baselineskip}
\setlength\belowcaptionskip{0.3\baselineskip}
	\centering
        \begin{minipage}{0.26\linewidth}
		\centering
        \begin{subfigure}{1\linewidth}
		\includegraphics[width=0.995\linewidth]{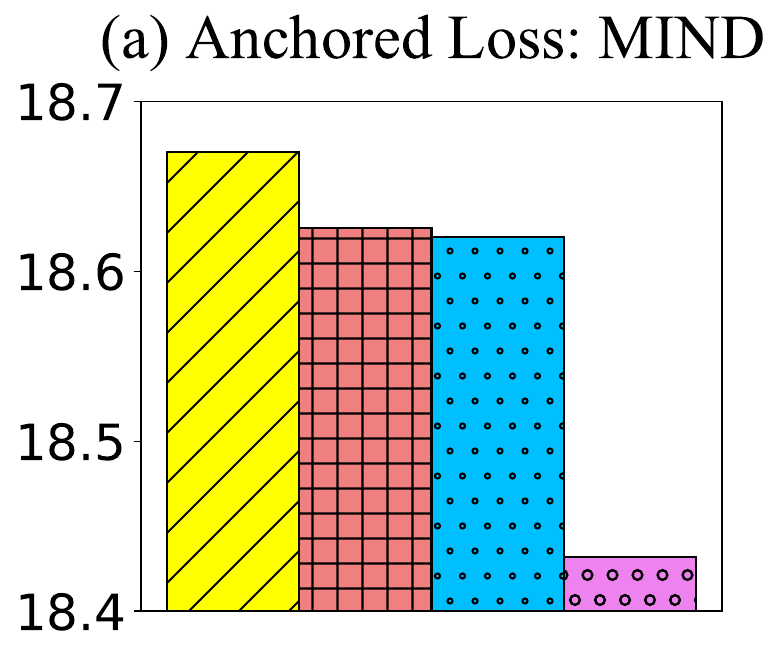}
		\label{fig:RQ3-4}
        \end{subfigure}
	\end{minipage}
	\begin{minipage}{0.27\linewidth}
		\centering
        \begin{subfigure}{1\linewidth}
		\includegraphics[width=0.995\linewidth]{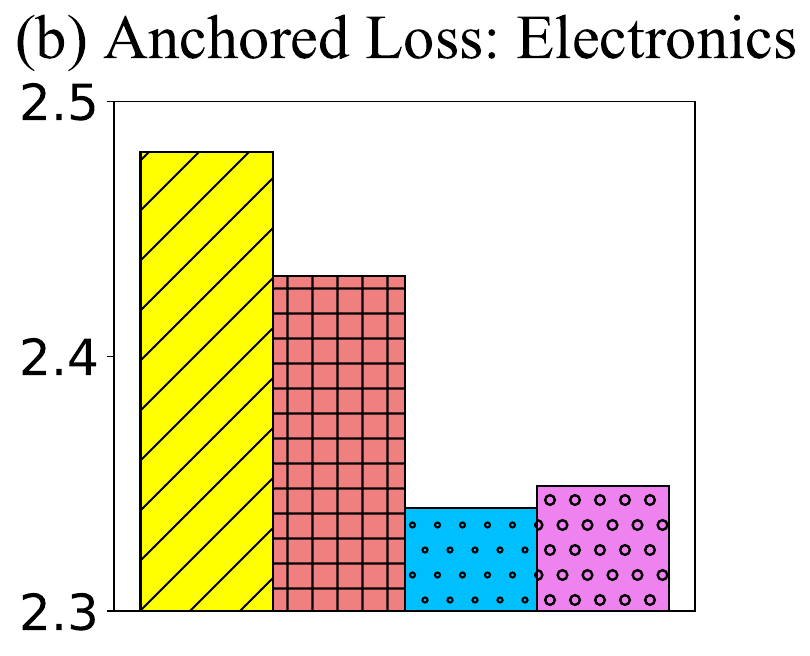}
		\label{fig:RQ3-1}
        \end{subfigure}
	\end{minipage}
	\begin{minipage}{0.445\linewidth}
		\centering
        \begin{subfigure}{1\linewidth}
		\includegraphics[width=0.995\linewidth]{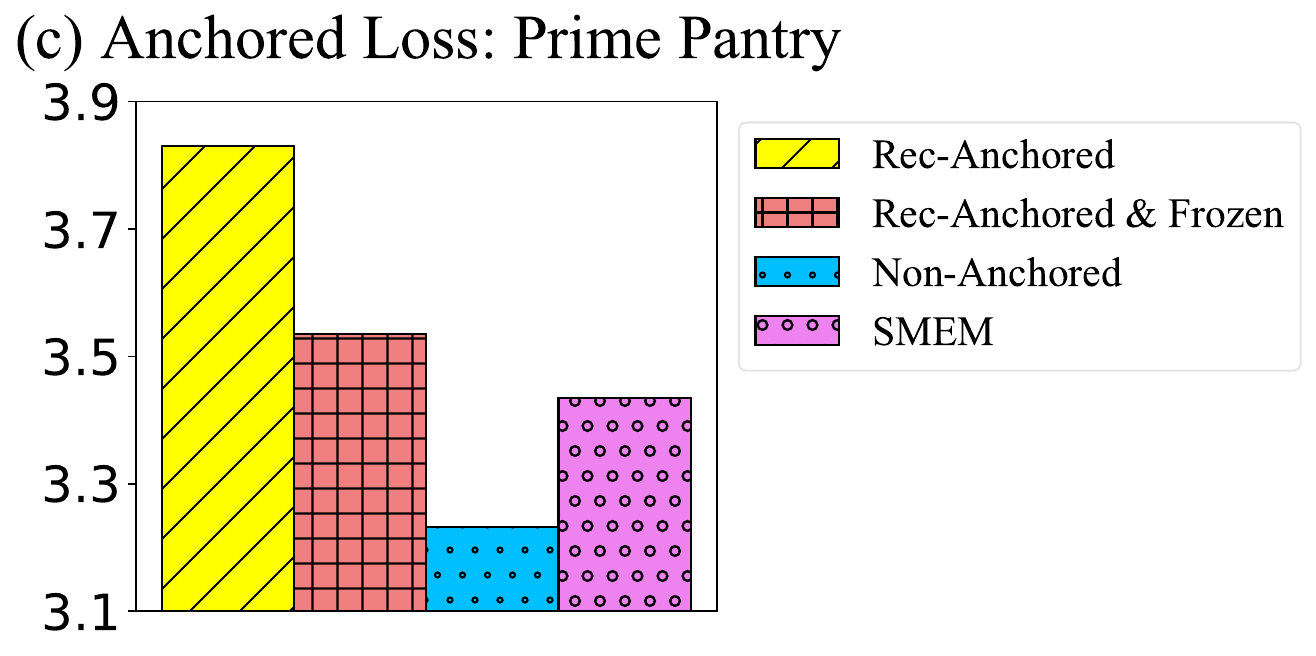}
		\label{fig:RQ3-2}
        \end{subfigure}
	\end{minipage}
	\caption{Comparison of anchored and non-anchored alignment losses on the Mind (left), Electronics (mid) and Prime Pantry (right) datasets. The y-axis denotes the HR@10.} 
	\label{fig:align}
\end{figure}

\subsection{Rec-Anchored Alignment Loss (RQ2)} \label{subsec:exp_rec_anchor}


We investigate the effect of recommendation anchoring in avoiding catastrophic forgetting by comparing the following three alignment losses in the align phase, namely: 
(1) \textbf{No Alignment}, \emph{i.e.}, \textbf{SMEM}, which doesn't involve any alignment loss but simply concat the textual and collaborative embeddings within the alignment expert,
(2) \textbf{Non-Anchored Alignment}, with only the alignment loss, \emph{i.e.}, $ D_k^2\left(\mathcal{D}_\text{text}, {\mathcal{D}_\text{rec}}\right)$,  
(3) \textbf{Rec-Anchored Alignment}, which combines the alignment loss with a BCE loss: $\mathcal{L}_\text{rec} + \gamma \cdot D_k^2\left(\mathcal{D}_\text{text}, \mathcal{D}_\text{rec}\right)$ and updates both collaborative and text embedding, and
(4) \textbf{Rec-Anchored and Frozen Alignment}, which freezes the collaborative embeddings upon the Rec-Anchored Alignment: $\mathcal{L}_\text{rec} + \gamma \cdot D_k^2\left(\mathcal{D}_\text{text}, \texttt{SG}(\mathcal{D}_\text{rec}) \right)$, where $\texttt{SG}$ denotes stop gradient operation. 
This can totally avoid catastrophic forgetting of the collaborative embeddings since they are frozen. 
However, this may hurt the alignment of the textual embeddings towards the collaborative space.

The results are shown in Fig.~\ref{fig:align}, and we can observe that: 
1) Our proposed Rec-Anchored Alignment performs the best among all losses;
2) Rec-Anchored losses, including Rec-Anchored as well as Rec-Anchored and Frozen, perform in general better than those non-anchored method, indicating that recommendation anchoring can effectively avoid catastrophic forgetting;
3) The Non-Anchored Alignment sometimes performs worse than Non-Alignment, indicating that simple alignment without anchoring even hurt the performance;
4) Rec-Anchored Alignment performs better than Rec-Anchored and Frozen Alignment, indicating the brute force of freezing collaborative embeddings hurts the alignment of the textual embeddings;

We further illustrate such forgetting of Non-Anchored as the violet lines in Fig.~\ref{fig:RQ3}(a) with the tool introduced in Sec.~\ref{subsec:discrepancy_measurement}, which show those cold item IDs (\emph{i.e.}, with bucket 7-10) has their embedding distance distribution more diverged from the original embeddings.

\begin{tcolorbox}[colback=blue!2!white,leftrule=2.5mm,size=title]
    \emph{Result 2. Our proposed Rec-Anchored Alignment loss avoids the catastrophic forgetting of collaborative embeddings, and succeeds in aligning textual embedding towards the collaborative space, achieving the best performance.}
\end{tcolorbox}

\begin{figure}[t]
\setlength\abovecaptionskip{-0.8\baselineskip}
\setlength\belowcaptionskip{0.2\baselineskip}
	\centering
	\begin{minipage}{0.46\linewidth}
		\centering
        \begin{subfigure}{1\linewidth}
		\includegraphics[width=0.995\linewidth]{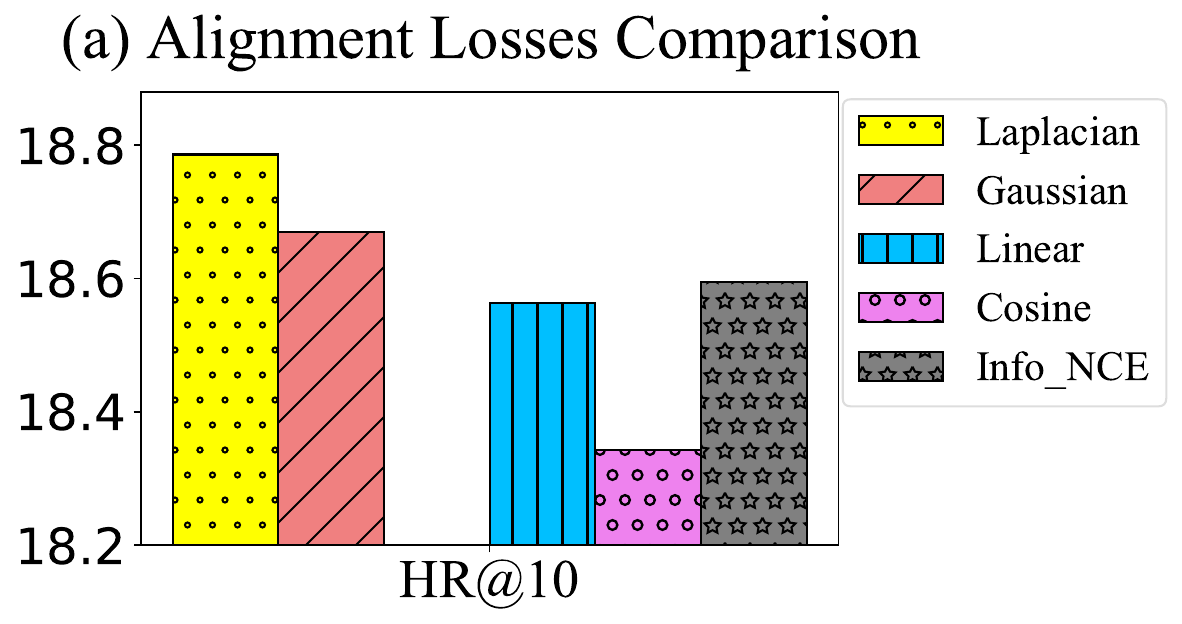}
		\label{fig:RQ4}
        \end{subfigure}
	\end{minipage}
	\begin{minipage}{0.53\linewidth}
		\centering
        \begin{subfigure}{1\linewidth}
		\includegraphics[width=0.995\linewidth]{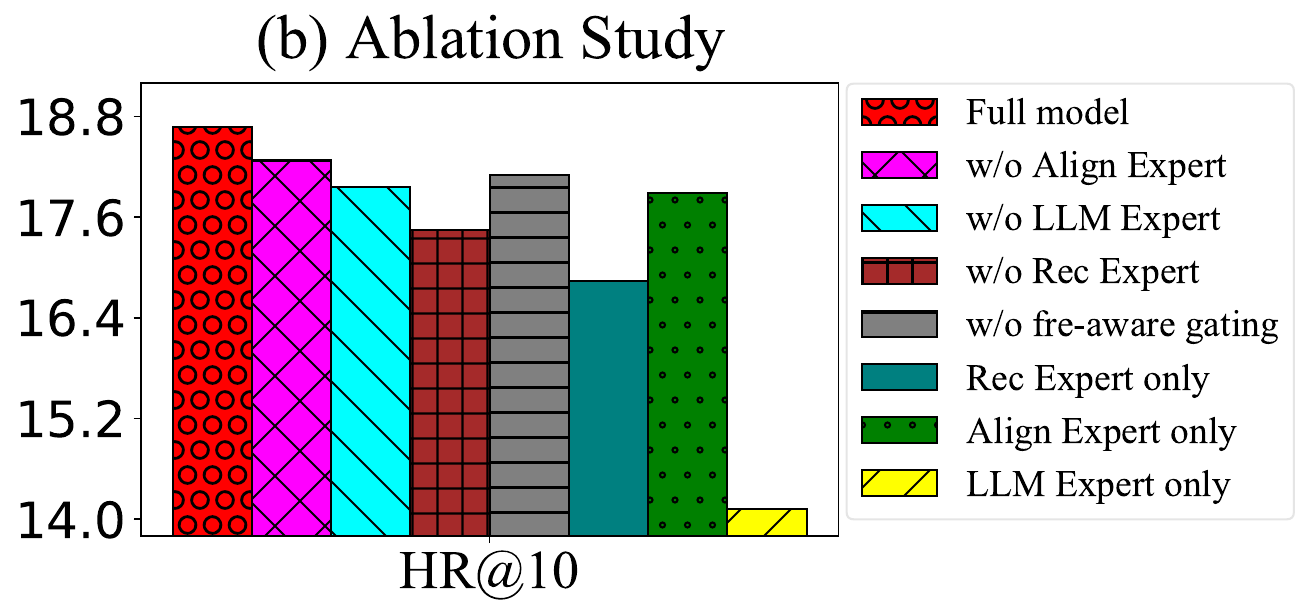}
		\label{fig:RQ5}
        \end{subfigure}
	\end{minipage}
	\caption{(a) Comparison of alignment losses and (b) ablation study comparing different model variants on MIND dataset.} 
	\label{fig:RQ4-6}
\end{figure}






\subsection{Characteristic Alignment Loss (RQ3)} \label{subsec:exp_characteristic}
To verify the advantages of characteristic kernels over non-characteristic ones in maximum mean discrepancy, apart from Gaussian kernels, we also experiment on the following three kernels and Info\_NCE loss (usually adopted by contrastive learning), in which only Laplacian kernels are characteristic on $\mathbb{R}^{d}$.

\begin{itemize}[leftmargin=*]
\item \textbf{Linear kernel}
\begin{equation}
    k(x,x') = \left \langle x,x'\right \rangle = x^\top x'
\end{equation}


\item \textbf{Cosine kernel}
\begin{equation}
    k(x,x') = 1- \left \langle \frac{x}{\|x\|}, \frac{x'}{\|x'\|} \right \rangle
\end{equation}

\item \textbf{Info\_NCE}
\begin{equation}
    \setlength{\abovedisplayskip}{3pt}
    \setlength{\belowdisplayskip}{1pt}
    \mathcal{L}_{CL}=-\sum_{i=1}^{n} \log \frac{\exp \left(k\left(\mathbf{x}_i, \mathbf{x}_i'\right) / \tau\right)}{\exp \left(k\left(\mathbf{x}_i, \mathbf{x}_i'\right) / \tau \right)+\sum_{j \neq i} \exp \left(k\left(\mathbf{x}_i, \mathbf{x}_j'\right) / \tau \right) }.
\end{equation} 
where $\tau$ is the temperature coefficient and cosine kernel is usually adopted to measure the sample similarity.

\item \textbf{Laplacian kernel}
\begin{equation}
    k(x,x') = \text{exp} ( -\frac{\|x-x'\|_1}{\sigma^2})
\end{equation}
\end{itemize}

Their results are shown in Fig.~\ref{fig:RQ4-6}(a), and we can see that characteristic kernels (\emph{i.e.}, Laplacian, and Gaussian) indeed perform better than non-characteristic ones. Notably, even if Info\_NCE explicitly pulls positive samples closer and negative samples farther thus achieving better results than pure cosine kernel, it is still inferior to characteristic kernels because the mean embedding of its RKHS is not rich enough. 
Therefore, we conclude that:

\begin{tcolorbox}[colback=blue!2!white,leftrule=2.5mm,size=title]
    \emph{Result 3. Characteristic kernels perform better than non-characteristic kernels in aligning textual embeddings towards collaborative space by maximum mean discrepancy.}
\end{tcolorbox}



\subsection{Towards Triple-Experts Architecture (RQ4)} \label{subsec:single_vs_multi}

\subsubsection{\textbf{Catastrophic Forgetting of Alignment}} \label{subsec:catastrophic_forgetting}

\begin{figure*}[t]
\setlength\abovecaptionskip{-1\baselineskip}
\setlength\belowcaptionskip{-0.2\baselineskip}
	\centering
        \begin{minipage}{0.24\linewidth}
		\centering
        \begin{subfigure}{1\linewidth}
		\includegraphics[width=0.995\linewidth,trim=0 20 0 0,clip]{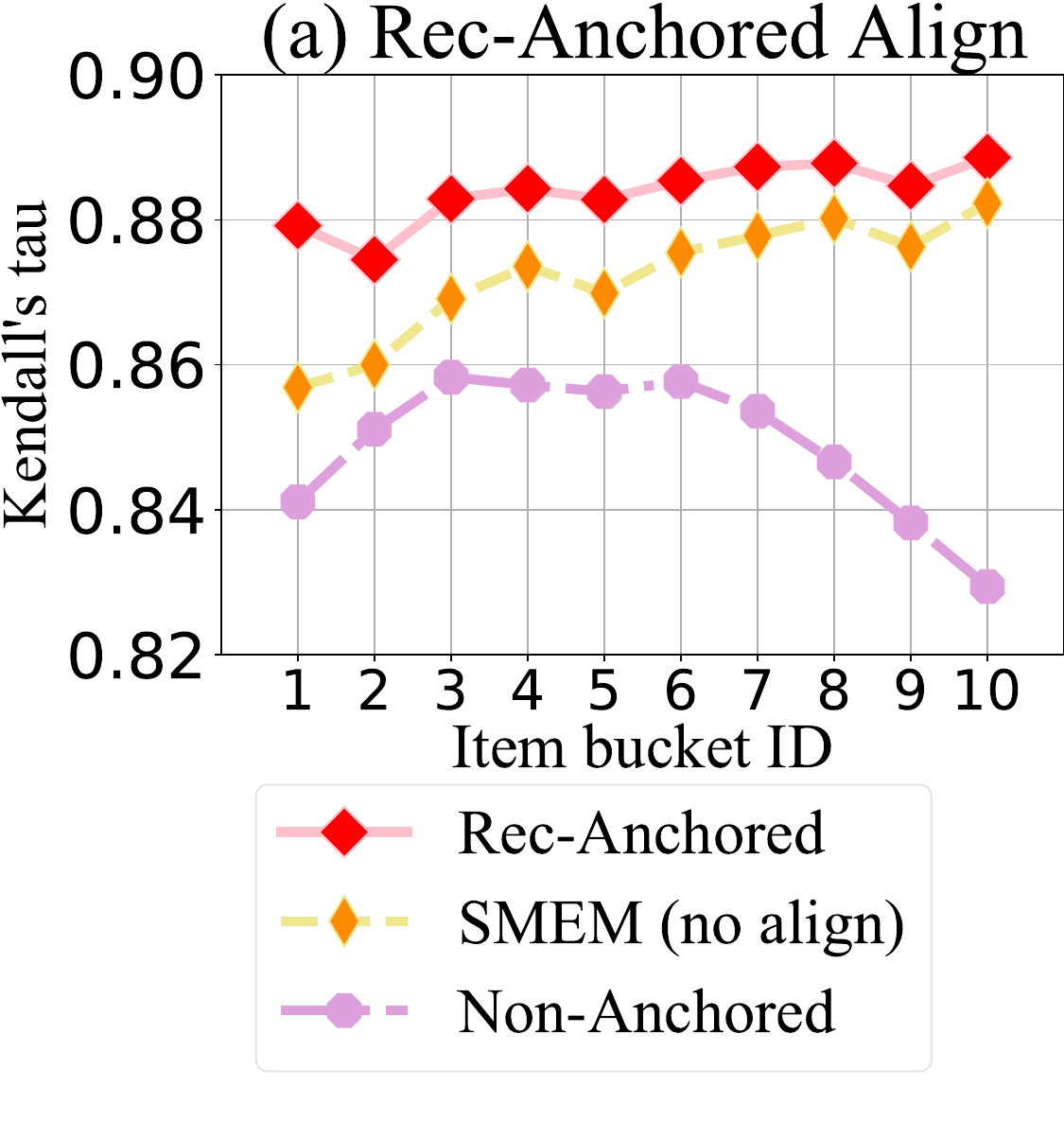}
		\label{fig:RQ3-4}
        \end{subfigure}
	\end{minipage}
	\begin{minipage}{0.24\linewidth}
		\centering
        \begin{subfigure}{1\linewidth}
		\includegraphics[width=0.995\linewidth,trim=0 35 0 0,clip]{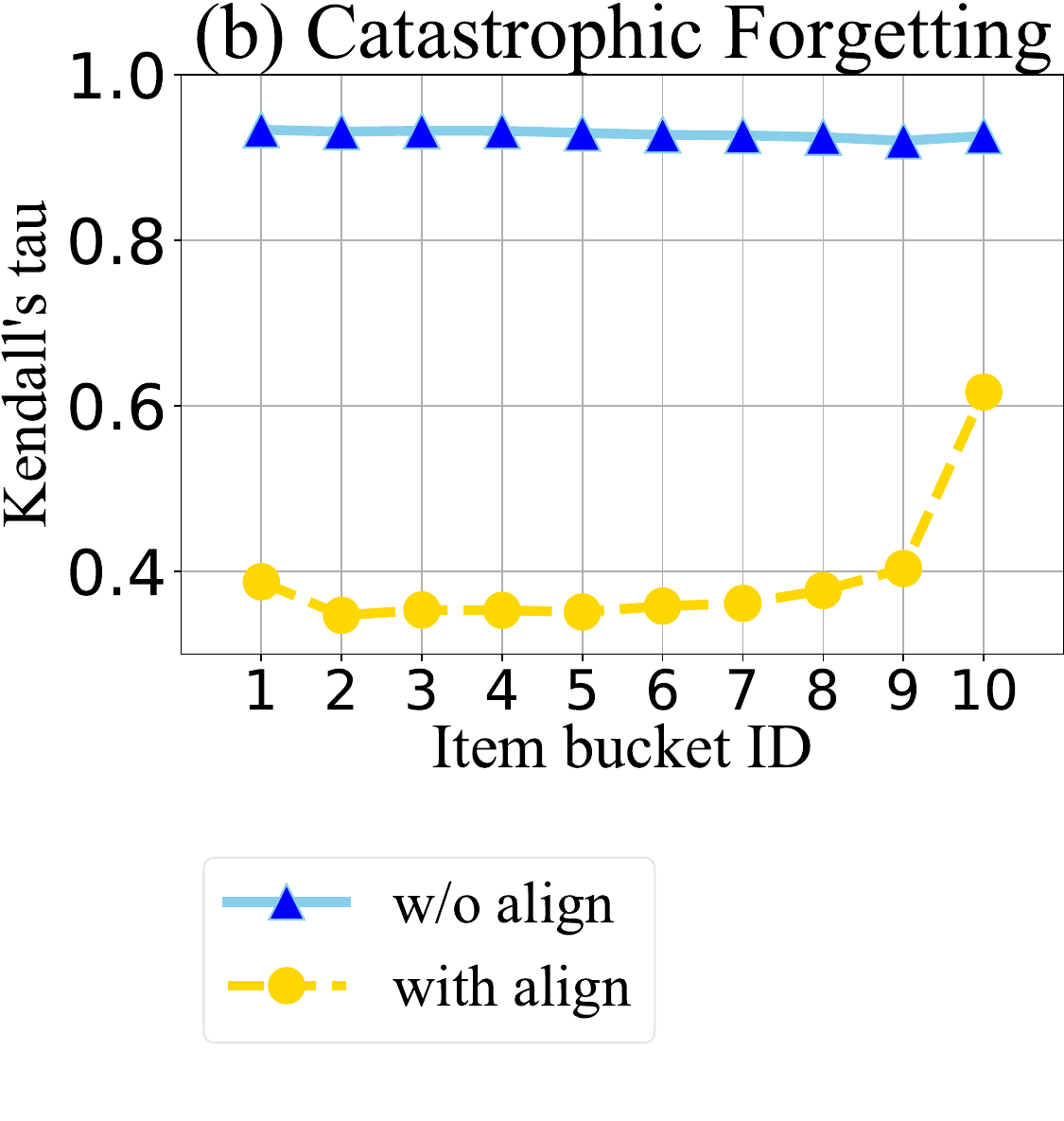}
		\label{fig:RQ3-1}
        \end{subfigure}
	\end{minipage}
	\begin{minipage}{0.24\linewidth}
		\centering
        \begin{subfigure}{1\linewidth}
		\includegraphics[width=0.995\linewidth,trim=0 20 0 0,clip]{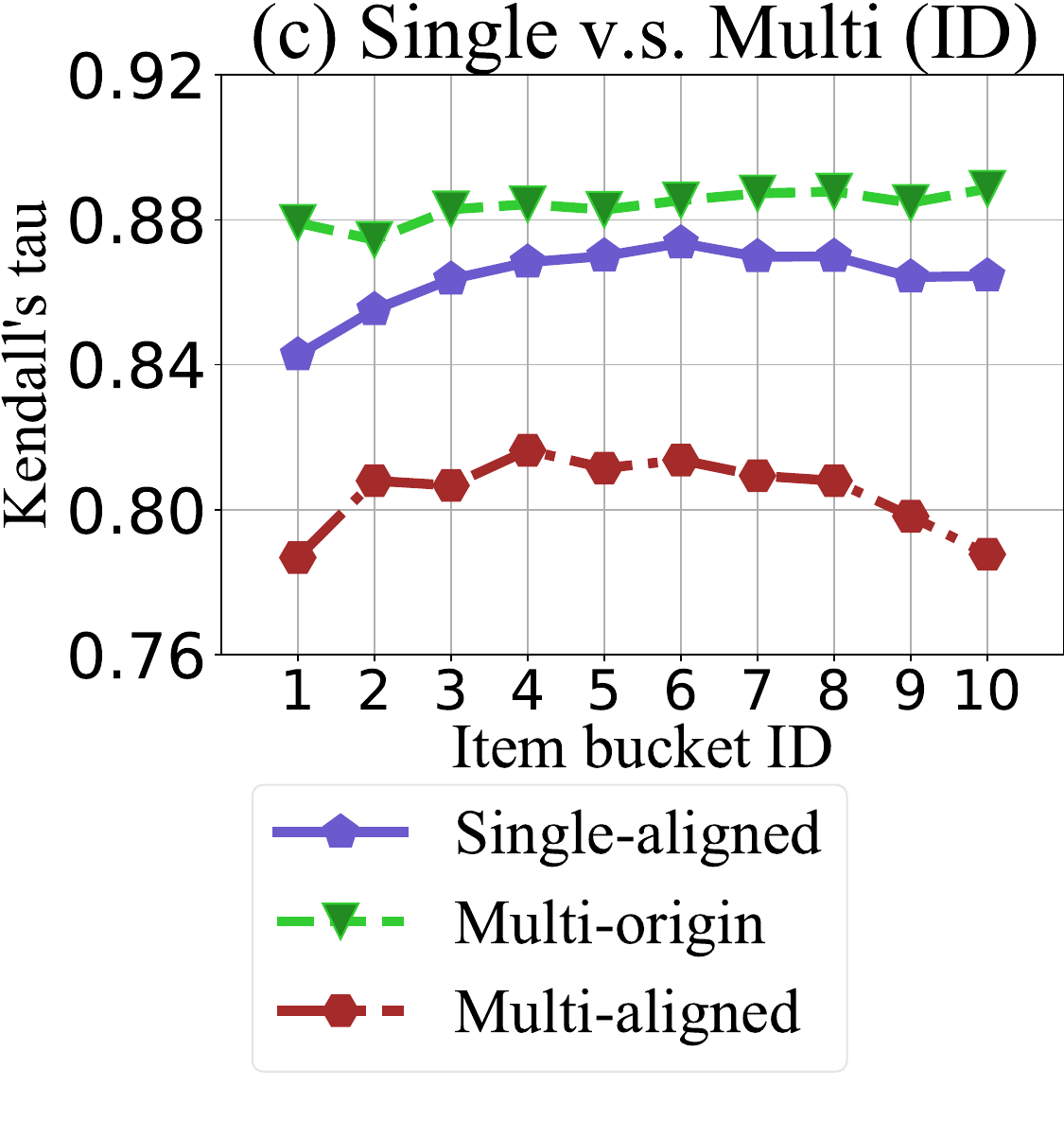}
		\label{fig:RQ3-2}
        \end{subfigure}
	\end{minipage}
        \begin{minipage}{0.24\linewidth}
		\centering
        \begin{subfigure}{1\linewidth}
		\includegraphics[width=0.995\linewidth,trim=0 20 0 0,clip]{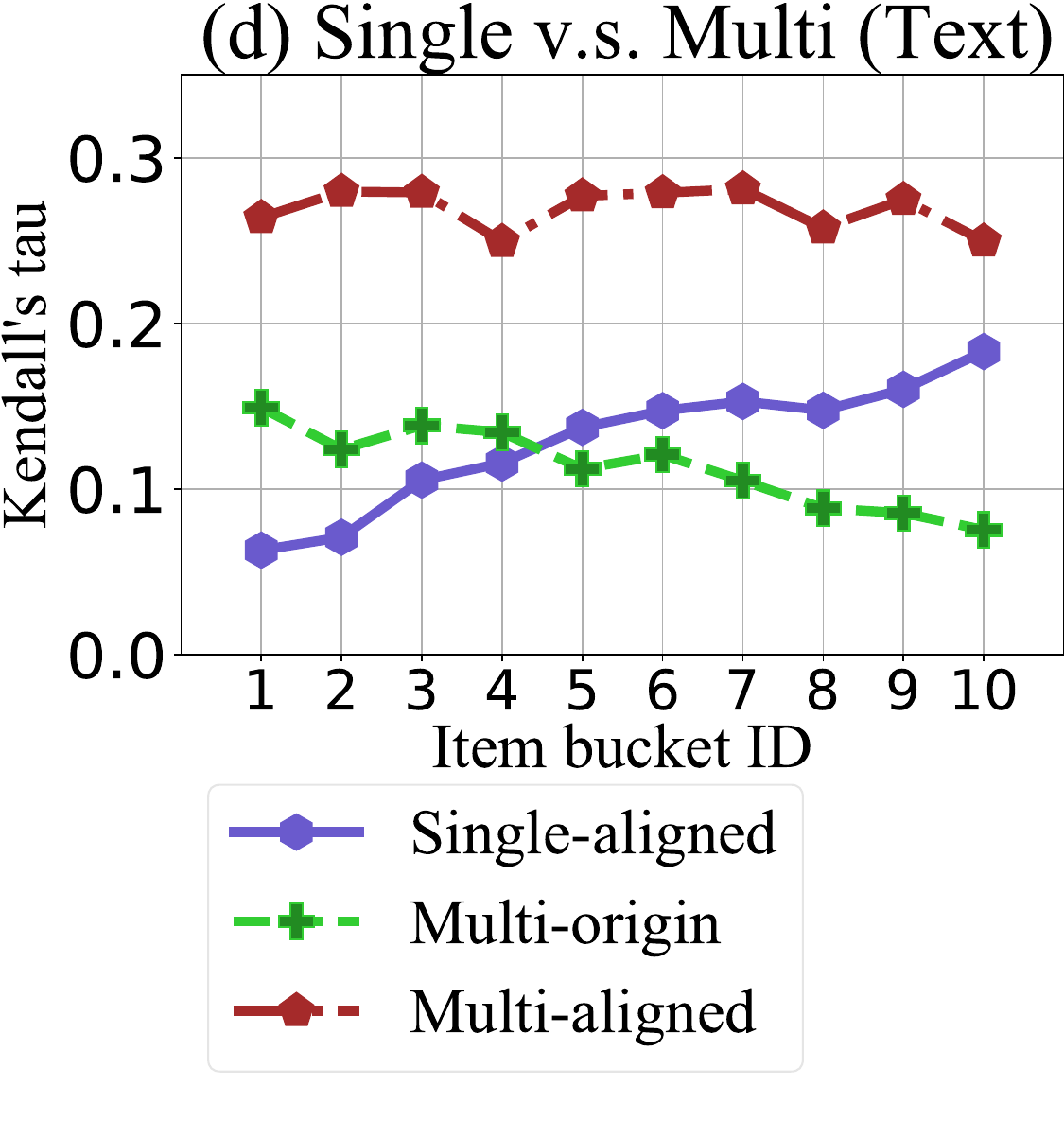}
		\label{fig:RQ3-2}
        \end{subfigure}
	\end{minipage}
	\caption{Kendall's tau between the Euclidean distance (same as L2 distance) of (a) collaborative embedding of different anchored alignment losses (b)  collaborative embedding of single-expert structure with or without alignment. (c)  collaborative embedding of single-expert and triple-experts (d) text embedding of single-expert and triple-experts. 
    They all use SASRec as backbone on MIND dataset. Item bucket ID from 1 to 10 denotes warm to cold.} 
	\label{fig:RQ3}
\end{figure*}

Many existing LLM4Rec methods~\cite{li2023ctrl} first align the collaborative embeddings of recommendation with the LLM, and then fine-tune these embeddings in the downstream recommendation task.
However, we wonder whether there is catastrophic forgetting of the alignment on the collaborative embeddings. 
That is, when there is only one single recommendation embedding table, whether the alignment leads to catastrophic damage to these embeddings, which \emph{CANNOT be recovered by the supervised fine-tuning}.

To validate this, we implement three methods, which all have only one single collaborative embedding table but differ in: 
the first method only employs the Binary Cross Entropy (BCE) loss upon the collaborative signal for supervised learning, without any alignment with the LLM, named SE-BCE in short; 
the second method employs BCE loss for both pre-training and fine-tuning, named SE-BCE-Tune; 
and the third method employs both BCE loss and the alignment loss for pre-training, then conducts a supervised fine-tuning using the collaborative labels, named SE-BCE-Align-Tune.

We measure the Kendall's tau of (SE-BCE, SE-BCE-Tune) and (SE-BCE, SE-BCE-Align-Tune) on the collaborative embedding distance of the methods and the results are shown in Fig.~\ref{fig:RQ3}(b). We divide the items into 10 buckets based on their frequency in the training set. 
The bucket ID range from 1 to 10, denoting from warm (high frequency) to cold (low frequency) items.
We observe that  SE-BCE-Align-Tune (\emph{i.e.}, the `with align' line in Fig.~\ref{fig:RQ3}(b)) has a much smaller value than the SE-BCE-Tune (\emph{i.e.}, the `w/o align' line in Fig.~\ref{fig:RQ3}(b)) on all buckets.
In particular, Kendall's tau on buckets 1 to 9 is less than 0.5, indicating that more than 50\% of the partial orders are disturbed.
Only the 10-th bucket gets Kendall's tau of 0.77.

\begin{tcolorbox}[colback=blue!2!white,leftrule=2.5mm,size=title]
    \emph{Result 4. With only one set of collaborative embeddings, the alignment loss leads to catastrophic forgetting on the collaborative embeddings.}
\end{tcolorbox}

The catastrophic forgetting of collaborative embeddings makes it hard to rely only on the aligned collaborative embeddings in the downstream recommendation tasks. 
To this end, we propose to incorporate two modality-specific experts in addition to the alignment expert, leading to a Triple-Experts architecture bellow.

\subsubsection{\textbf{Triple-Experts Architecture}}

To illustrate the effect and disentanglement capability of triple-experts architecture, on MIND dataset and SASRec as the backbone model, we implement SE-BCE, SE-BCE-Align-Tune, and TE-BCE-Align-Tune whose training objective is the same as SE-BCE-Align-Tune but implemented on triple-experts with two embedding tables for both ID and text embedding. 
We measure the Kendall's tau of (SE-BCE, SE-BCE-Align-Tune) (\emph{i.e.}, the `Single-aligned' line in Fig.~\ref{fig:RQ3}(c) and (d)) and (SE-BCE, TE-BCE-Align-Tune) (\emph{i.e.}, the `Multi-origin' and `Multi-aligned' line in Fig.~\ref{fig:RQ3}(c) and (d)). The results of ID and text embedding are shown in Fig.~\ref{fig:RQ3}(c) and (d), where `Single-aligned' denotes the ID or text embedding of SE-BCE-Align-Tune. `Multi-origin' and `Multi-aligned' denote two sets of ID or text embeddings of triple-experts. We have the following observations.

First, the Kendall's tau between the distance of pre-trained and updated collaborative embedding is generally larger than that between the distance of pre-trained and updated text embedding. This is because the text data act as auxiliary information of recommendation, thus the aligned space is more closer to the original collaborative space than the semantic space.
Second, the Kendall's tau of at least one set of embedding in triple-experts (yellow and purple line) is consistently larger than that of embedding in single-expert structure (red line) on all item buckets. 
Third, TE-BCE-Align-Tune surpasses SE-BCE-Align-Tune by 3.59\% on nDCG@10 on MIND dataset.
Therefore, we can conclude that too much deviation of the aligned collaborative embedding distribution from the original one has negative effect on the recommendation performance. A possible reason is some unique information (\emph{e.g.}, the rank of distance between all behavior and item embedding pairs) in the original embedding distribution is lost.
These results indicates that compared with single-expert, triple-experts architecture could retain more information of the original embedding, thus being capable of alleviating the catastrophic forgetting and plays a vital role of disentanglement leading to performance enhancement.

\subsection{Ablation Study (RQ5)}\label{subsec:abalation}

We conduct an ablation study on the MIND dataset to investigate the effect of each expert and the frequency-aware gating. 
The results are shown in Fig.~\ref{fig:RQ4-6}(b).
First, we remove each expert in the last phase and observe the performance deteriorates when we remove any of the three experts.
Interestingly, removing the recommendation (ID) expert leads to the largest performance drop.
This indicates that the alignment loss leads to inevitable information loss of the collaborative embeddings, even with our Rec-Anchored loss.
Therefore, the recommendation-specific tower is essential to retain the collaborative signals.

Second, we try to retain only one of these three expert and compare their performance.
Surprisingly, the single alignment model achieves the best performance.
This is possibly due to the fact that, on the one hand, it keeps the collaborative semantics through the Rec-Anchored alignment loss, and on the other hand, it boosts the performance on the cold items with the help of LLM semantics.




Finally, the model variant removing the frequency-aware gating fuses all experts with the learned weights on all items equally, leading to a decrease of 3.06\% in HR@10. 
Therefore, from the perspective of target item frequency, this gating acts as an bridge connecting the collaborative, semantic, and alignment space adaptively considering their different credibility.


\begin{table*}[t]
\centering
\belowrulesep=0pt
\aboverulesep=0pt
\setlength\abovecaptionskip{0.2\baselineskip}
\setlength\belowcaptionskip{0.2\baselineskip}
\caption{Compatibility experiments on MIND, Electronics, and Prime Pantry dataset. `Origin' denotes the model trained on ID only and `Impr.' indicates the improvement against the original backbone model.}
\label{tab:comp}

\resizebox{0.995\textwidth}{!}{
\begin{tabular}{@{}c|cccccc|cccccc@{}}

\toprule
\multirow{3}{*}{{Datasets}} & \multicolumn{6}{c|}{GRU4Rec} & \multicolumn{6}{c}{Caser} \\ \cmidrule(l){2-13}
 & \multicolumn{2}{c}{Origin} & \multicolumn{2}{c}{+\name} & \multicolumn{2}{c|}{{Impr.}} & \multicolumn{2}{c}{Origin} & \multicolumn{2}{c}{+\name} & \multicolumn{2}{c}{{Impr.}} \\
 & HR@10 & nDCG@10 & HR@10 & nDCG@10 & HR@10 & nDCG@10 & HR@10 & nDCG@10 & HR@10 & nDCG@10 & HR@10 & nDCG@10 \\ \midrule
 MIND & 13.3969 & 6.9098 & 16.5762 & 8.8475 & 23.73\% & 28.04\%  & 8.6879 & 3.7981 & 9.0569 & 4.2696 & 4.25\% & 12.41\%   \\
 Electronics & 0.8437 & 0.4104 & 1.1016 & 0.5898 & 30.57\% & 43.73\% & 0.5746 & 0.2355 &  0.6351 & 0.2501 & 10.53\% & 6.20\% \\
 Prime Pantry & 1.7248 & 0.8227 & 2.5484 & 1.2307 & 47.75\% & 49.59\% &  0.6526  &  0.4136 & 1.4063 & 0.5507 & 115.49\% & 33.16\%  \\

\bottomrule

\end{tabular}}
\end{table*}

\subsection{LLM2Vec v.s. BERT (RQ6)}
To compare the performance of LLM and pre-train language model (PLM) like BERT, we replace LLM2Vec with BERT-Base as the text encoder of items. It achieves the performance of HR@10~= 18.3989 and nDCG@10~= 9.9810 on MIND, which is inferior to LLM2Vec achieving HR@10~= 18.6703 and nDCG@10~= 10.1515. We speculate the reasons are: 1) More parameters of Llama3-8B bring in more powerful encoding capability than BERT-Base with 110 million parameters only. 2) Compared with BERT, LLM2Vec is trained with more abundant text data on wiki corpus and adopts a different Unsupervised contrastive training (SimCSE) objective.

\subsection{Compatibility (RQ7)}

To validate the compatibility of \name, apart from SASRec~\cite{kang2018self}, we also experiment on two typical sequential recommendation models as backbones including GRU4Rec~\cite{hidasisession} and Caser~\cite{tang2018personalized}. As shown in Tab.~\ref{tab:comp}, \name significantly enhance the performance of the original model on all three datasets, indicating that it acts as a powerful model-agnostic enhancement paradigm.

%% file: 4.related_work.tex
\section{Related Work} \label{rw}
This section summarizes the related work on sequential recommendation, large language model, and multi-modal recommendation.

\subsection{Sequential Recommendation}

Sequential recommendation (SR) focuses on modeling the sequential dependency over the interaction sequence to capture user preference.
Inspired by the success of self-attention in natural language processing~\cite{vaswani2017attention}, SASRec~\cite{kang2018self} incorporates it into SR and
afterward many works explore target-attention recommendation in a similar manner~\cite{din2018, dien2019, dsin2019, tin2024, sim2020, si2024twin, chang2023twin, feng2024long}.
Caser~\cite{tang2018personalized} and GRU4Rec~\cite{hidasisession} adopts convolution and recurrent networks for sequential modeling.
Nevertheless, these models mainly takes ID as the only input thus suffering from the cold-start problem. By contrast, our proposed \name paradigm is capable of enhancing them as backbones by capturing complex correlation of different modality data through our designed alignment and triple-experts structure.

\subsection{Large Language Model \& Multi-Modal Rec}

In the recommendation community, some works explicitly adopt alignment between tabular and textual data which could achieve information gain. For example, CTRL~\cite{li2023ctrl} leverages contrastive learning, and FLIP~\cite{wang2023flip} uses data masking and reconstruction pre-training with contrastive learning. Afterward, some works like DisCo~\cite{du2024disco} and DaRec~\cite{yang2024darec} conduct disentanglement to capture the specific information. However, they simply separate the embeddings into different parts which may not be fully learned by the common single-expert structure. By contrast, our \name adopts triple-experts in a frequency-aware manner to adaptively fuse different information, which achieves better disentanglement ability.

Meanwhile, numerous studies~\cite{liu2019user, sun2019bert4rec, he2016vbpr, chen2017attentive, chen2019personalized, rajput2024recommender, singh2024better, scalable2024zhang,liang2023mmmlp} focused on enhancing recommender systems by incorporating multi-modal content. 
Recently, researchers have explored capturing user fine-grained preferences across different modalities. 
For instance, MMGCN~\cite{wei2019mmgcn} attempts to model user preferences using modal-specific user-item bipartite graphs. 
M3SRec~\cite{bian2023multi} proposes a novel multi-modal mixture-of-experts (MoE) fusion network. Taobao
~\cite{sheng2024enhancing} proposes a two-stage paradigm to enhance recommendation with multi-modality recommendation.
M3CSR ~\cite{chen2024multi} proposes a multi-modality framework for to tackle cold-start in short-video recommendation.
However, the existing methods either use a non-characteristic alignment loss or don't employ a disentangled expert, while our method has both of these two critical parts.



%% file: 5.conslusion.tex
\section{Conclusion} \label{conc}

In this paper, we propose a novel Pre-train, Align, and Disentangle (PAD) paradigm to empower sequential recommendation with large language models. 
We propose a characteristic recommendation-anchored loss to better align textual embeddings towards the collaborative space and avoid catastrophic forgetting.
We employ a triple-experts architecture, consisting of aligned and modality-specific experts with disentangled embeddings, which is fine-tuned in a frequency-aware manner.
Comprehensive experiments on three public datasets validate the effectiveness of our proposed method. Notably, the framework can be extended to multi-modal modeling.



%% file: 6.appendix.tex
\section*{ACKNOWLEDGEMENT}
This research was partially supported by Tencent Rhino-Bird Focused Research Program, Research Impact Fund (No.R1015-23), and Collaborative Research Fund (No.C1043-24GF).

\section{Experimental Settings} \label{detail}

\begin{figure*}[t]
\setlength\abovecaptionskip{-1\baselineskip}
\setlength\belowcaptionskip{-0.2\baselineskip}
	\centering
        \begin{minipage}{0.47\linewidth}
		\centering
        \begin{subfigure}{1\linewidth}
		\includegraphics[width=0.995\linewidth]{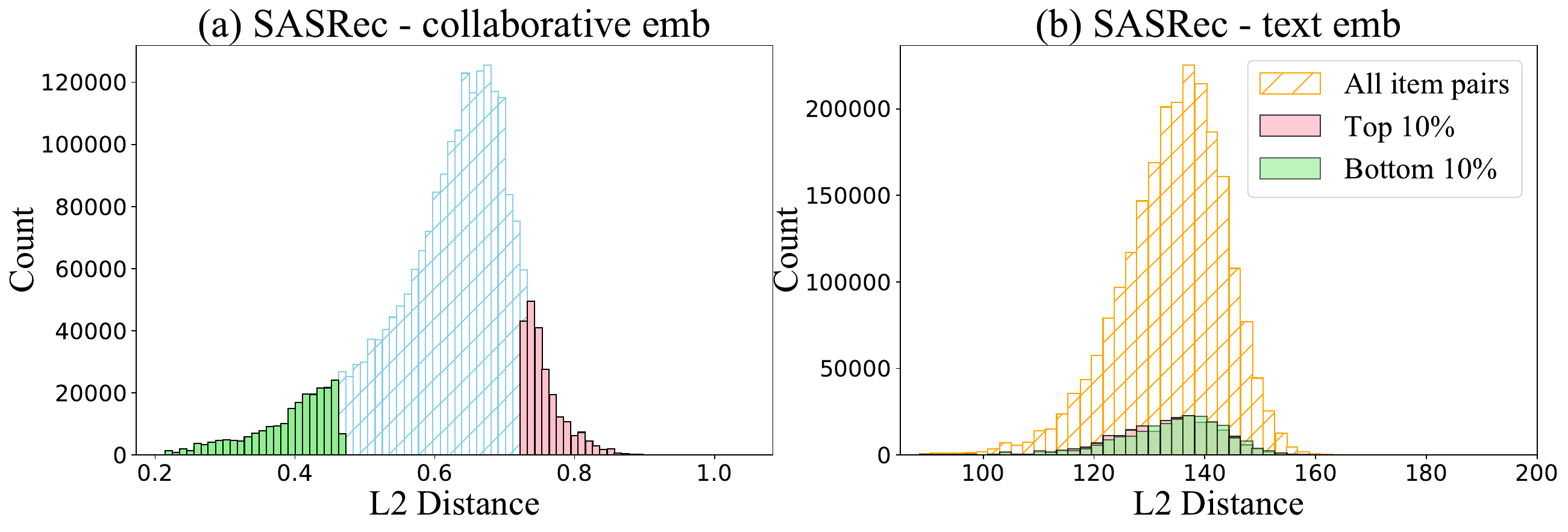}
		\label{fig:RQ3-4}
        \end{subfigure}
	\end{minipage}
        \begin{minipage}{0.47\linewidth}
		\centering
        \begin{subfigure}{1\linewidth}
		\includegraphics[width=0.995\linewidth]{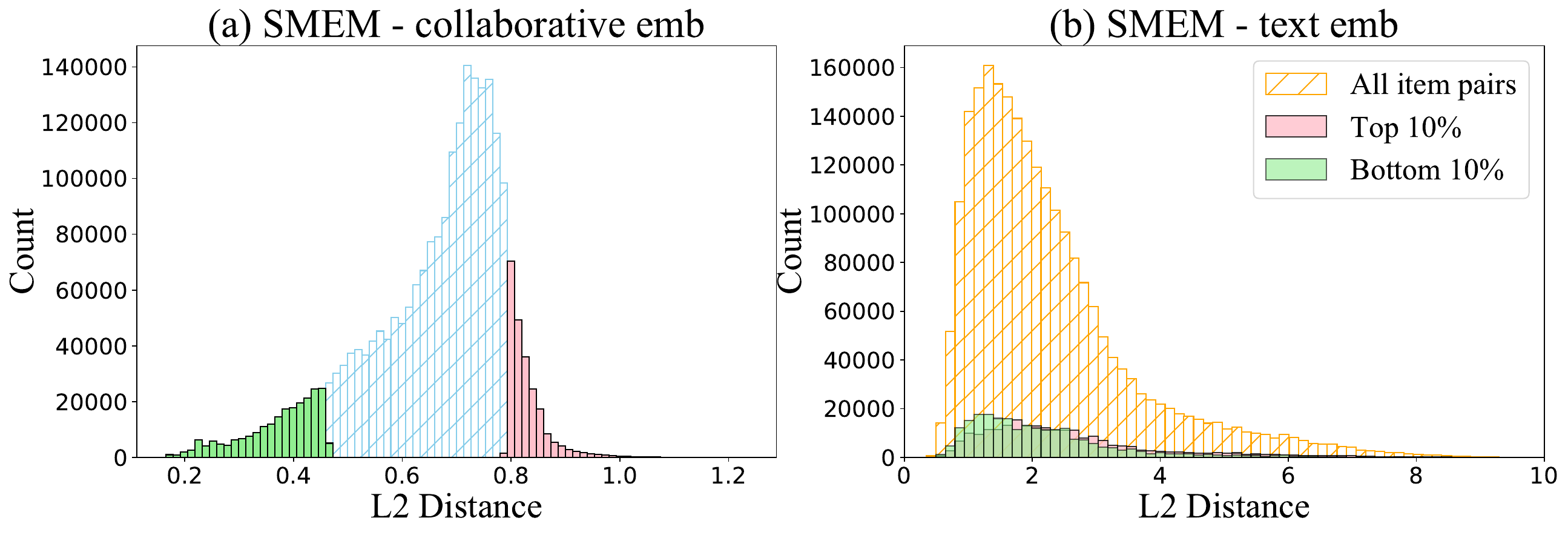}
		\label{fig:RQ3-4}
        \end{subfigure}
	\end{minipage}

        \vspace{-2mm}
 
        \begin{minipage}{0.47\linewidth}
		\centering
        \begin{subfigure}{1\linewidth}
		\includegraphics[width=0.995\linewidth]{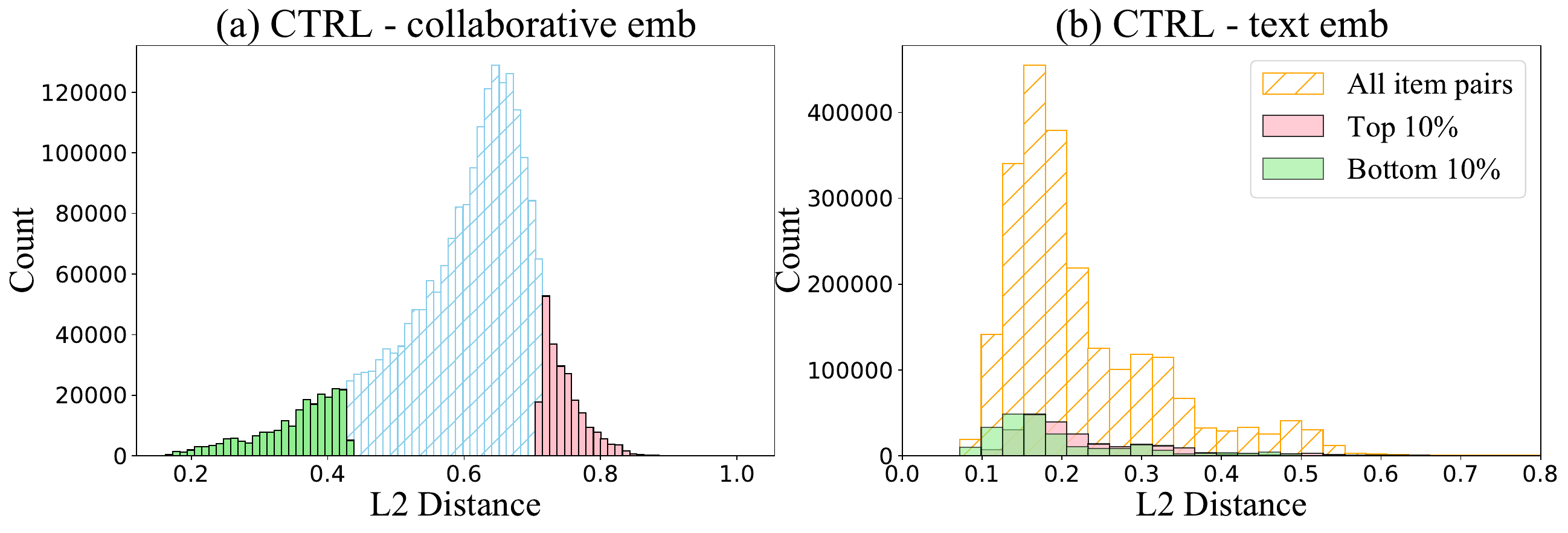}
		\label{fig:RQ3-4}
        \end{subfigure}
	\end{minipage}
        \begin{minipage}{0.47\linewidth}
		\centering
        \begin{subfigure}{1\linewidth}
		\includegraphics[width=0.995\linewidth]{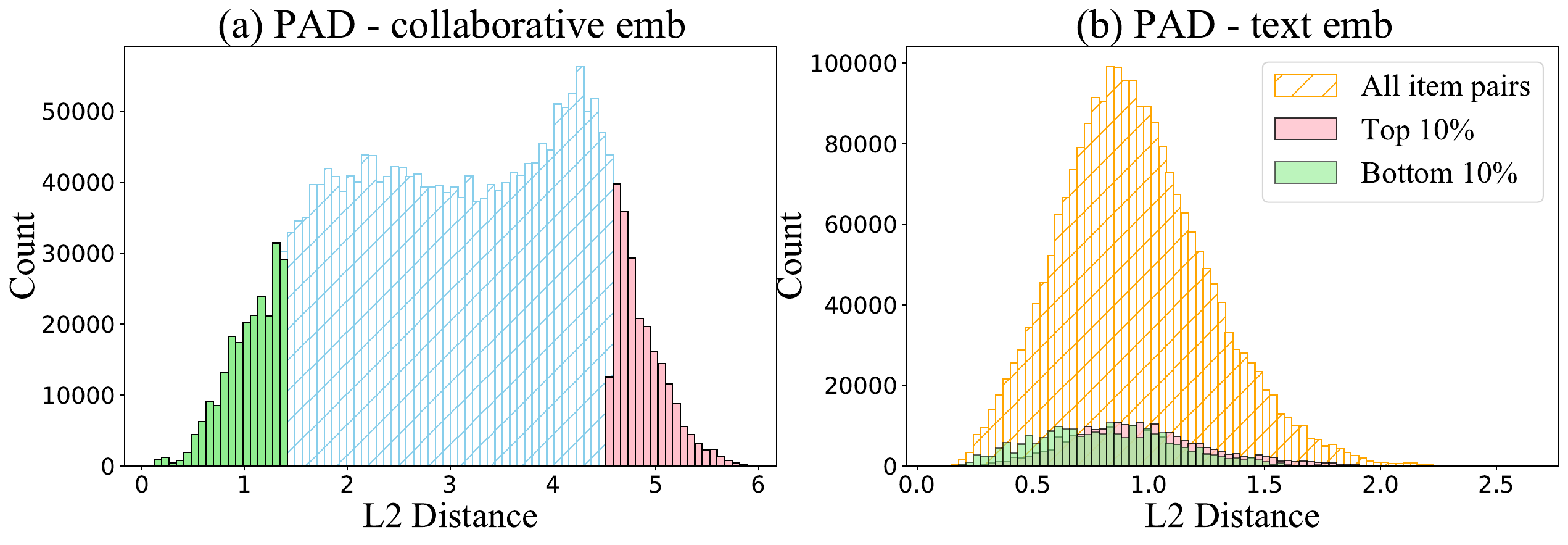}
		\label{fig:RQ3-4}
        \end{subfigure}
	\end{minipage}
	\caption{$\mathcal{P}_{\text{Top-10\% }}^\text{ID}$ and $\mathcal{P}_{\text{Bottom-10\% }}^\text{ID}$ under the distance distribution regarding the collaborative and text embeddings in the original SASRec, SMEM, CTRL, and \name on warm items.} 
	\label{app:warm}
\end{figure*}

For LLM2Vec we choose the Llama3-8B as the base model with the unsupervised-trained LoRA weights. We leverage the news title on MIND and item title on Amazon to generate text embedding. Suppose $l$ is defined as the length of user's historical interaction sequence. For all the three datasets, we select the users with no less than 5 interactions and extract the latest 23 items interacted. 
The task is to predict whether the user would click the $\left(l-2\right)$-th, $\left(l-1\right)$-th, and $l$-th item on training, validation, and test set, respectively, given the previous interacted items. Besides, AdamW~\cite{loshchilov2017fixing} is adopted as optimizer. Batch size is 16, the early stop epoch is set to 10, and dropout probability is 0.1. $L$2 regularization is adopted with weight of 0.1. The weight $\gamma$ is searched and set to 0.2.

\section{Pseudo-code}

\begin{algorithm}[t]

\caption{\name paradigm for sequential recommender systems (SRS)} \label{alg}

\KwIn{User set $\mathcal{U}$; item set $\mathcal{I}$; historical interaction sequence; item text embedding; true label $y$;}
\KwOut{A well trained SRS with three experts.}



\nonl Phase 1: Pre-train\

\While{not converge}{
 Sample a mini-batch data from $\mathcal{U}$\;
 Calculate the prediction result of sequential rec model\;
 Calculate the BCE loss\;  
 Take the gradient and update sequential rec model\; 
}

\nonl Phase 2: Rec-Anchored Alignment\

Load pre-trained ID embedding\

\While{not converge}{
 Sample a mini-batch data from $\mathcal{U}$\;
 Retrieve text embedding of behavior items\;
 Calculate the prediction result of alignment expert\;
 Calculate the BCE loss plus MMD\; 
 Take the gradient and update alignment expert\; 
}

\nonl Phase 3: Triple-Experts Fine-tune\

Load parameters of pre-trained Rec-specific \& Alignment expert \

\While{not converge}{
 Sample a mini-batch data from $\mathcal{U}$\;
 Retrieve text embedding of behavior items\;
 Calculate the prediction result of all three experts\;
 Fuse the results via frequency-aware gating\; 
 Calculate the BCE loss\;  
 Take the gradient and update all parameters\;
}

\end{algorithm}

The general procedure of \name is given in
Alg.~\ref{alg}, which consists of three phases: (1) LLM \& Recommendation Pre-train (from Line 1 to 6), (2) Alignment with MK-MMD (from Line 7 to 14), and (3) Recommendation Supervised Fine-tuning (from Line 15 to 23).

\section{Advanced Analysis}

Denote $\mathcal{P}_{\text{Top-10\% }}^\text{ID}$ and $\mathcal{P}_{\text{Bottom-10\% }}^\text{ID}$ as the group of item pairs with the top-10\% and bottom-10\% distance based on collaborative ID embeddings. 
Apart from the results on cold items in Fig.~\ref{app:cold}, we present the distribution of these item pairs under the distance distribution regarding the collaborative and textual embeddings learned by each model on warm items in Fig.~\ref{app:warm}. It indicates that the distribution of $\mathcal{P}_{\text{Top-10\% }}^\text{ID}$ and $\mathcal{P}_{\text{Bottom-10\% }}^\text{ID}$ can not be differentiated in textual modality probably because all these recommendation models mainly rely on collaborative information for warm items.